\pdfoutput=1

\documentclass[sigconf, nonacm, pdfa]{acmart}

\newcommand\vldbdoi{XX.XX/XXX.XX}
\newcommand\vldbpages{XXX-XXX} 
\newcommand\vldbvolume{19}
\newcommand\vldbissue{10}
\newcommand\vldbyear{2026}
\newcommand\vldbauthors{\authors}
\newcommand\vldbtitle{\shorttitle} 
\newcommand\vldbavailabilityurl{https://github.com/sjoon-oh/aker}

\newcommand\vldbpagestyle{empty} 

\usepackage{graphicx}       
\usepackage{caption}
\usepackage{subcaption}     
\usepackage{kotex}          
\usepackage{comment}        
\usepackage{enumitem}
\usepackage{adjustbox, multirow}
\usepackage{makecell}
\usepackage{multicol}

\usepackage[ruled,linesnumbered,noend]{algorithm2e}

\usepackage{hyperref}
\usepackage{ulem}

\usepackage[switch]{lineno}
\usepackage{lipsum}

\usepackage{listings}
\usepackage[dvipsnames]{xcolor}

\usepackage{booktabs}

\usepackage{amssymb}
\usepackage{pifont}

\usepackage{mathtools}
\usepackage{wasysym}
\usepackage{multirow}
\usepackage{geometry}
\usepackage[skins]{tcolorbox}
\usepackage{framed}

\usepackage{balance}
\usepackage{subfiles}

\usepackage[colorinlistoftodos]{todonotes}

\usepackage[a-2b]{pdfx}
\newcommand{\olive}[1]{\textcolor{olive}{#1}}

\usepackage{mdframed}
\usepackage{multirow}

\newcommand{\aker}{{\textsc{Aker}}}

\begin{document}


\title{
    \aker: Density-Aware Approximate Caching for Vector Search 
}
\subtitle{
    Extended Version
}



\settopmatter{authorsperrow=4}

\author{Sukjoon Oh}
\authornote{Work partially performed during the internship at Microsoft Research.}
\affiliation{%
  \institution{KAIST}
}
\email{sjoon@kaist.ac.kr}

\author{Minki Kang}
\affiliation{%
  \institution{KAIST}
}
\email{minki.kang@kaist.ac.kr}

\author{Dohyun Kim}
\affiliation{%
  \institution{KAIST}
}
\email{ehgus421210@kaist.ac.kr}

\author{Baotong Lu}
\authornote{Corresponding authors: Baotong Lu and Youjip Won.}
\affiliation{%
  \institution{Microsoft Research}
}
\email{baotonglu@microsoft.com}

\author{Jing Liu}
\affiliation{%
  \institution{Microsoft Research}
}
\email{jingliu3@microsoft.com}

\author{Qianxi Zhang}
\affiliation{%
  \institution{Microsoft Research}
}
\email{qiazh@microsoft.com}

\author{Qi Chen}
\affiliation{%
  \institution{Microsoft Research}
}
\email{cheqi@microsoft.com}

\author{Youjip Won}
\authornotemark[2]
\affiliation{%
  \institution{KAIST}
}
\email{ywon@kaist.ac.kr}

\begin{abstract}
    Disk-based approximate nearest neighbor search (ANNS) incurs high I/O overhead due to frequent disk accesses during index traversal. 
Approximate caching, which reuses the results of past queries to serve future similar queries, offers a promising approach to bypass expensive disk searches. 
However, existing approaches suffer from two key limitations. 
First, their approximate hit predicates fail to simultaneously achieve high throughput and high accuracy, 
as they do not adapt to the varying local neighbor density in high-dimensional spaces. 
Second, they lack an effective refresh mechanism to maintain cache correctness under vector updates.

We present \aker, an approximate cache for disk-based ANNS. \aker~addresses these limitations through two core design choices. 
First, we introduce a per-query similarity threshold, 
where each cache entry maintains its own threshold that is dynamically adjusted based on observed cache hit patterns. 
This design enables \aker~to adapt to neighborhood densities to preserve both efficiency and accuracy. 
Second, we propose \textit{del-consistency}, a consistency model for ANNS caches that applies deletions eagerly and insertions lazily. 
Under this model, \aker~ implements a low-overhead refresh mechanism that bounds cache staleness and preserves high search accuracy. 
We integrate \aker~ into pgvector and evaluate it on representative workloads. 
\aker~ improves recall by up to $64$ percentage points over prior solutions and increases QPS by up to $3.2\times$, while using $0.6\times$ the memory of pgvector's shared buffers.
\end{abstract}

\maketitle

\pagestyle{\vldbpagestyle}
\begingroup\small\noindent\raggedright\textbf{PVLDB Reference Format:}\\
\vldbauthors. \vldbtitle. PVLDB, \vldbvolume(\vldbissue): \vldbpages, \vldbyear.\\
\href{https://doi.org/\vldbdoi}{doi:\vldbdoi}
\endgroup
\begingroup
\renewcommand\thefootnote{}\footnote{\noindent
\noindent
This work is licensed under the Creative Commons BY-NC-ND 4.0 International License. Visit \url{https://creativecommons.org/licenses/by-nc-nd/4.0/} to view a copy of this license. For any use beyond those covered by this license, obtain permission by emailing \href{mailto:info@vldb.org}{info@vldb.org}. Copyright is held by the owner/author(s). Publication rights licensed to the VLDB Endowment. \\
\raggedright Proceedings of the VLDB Endowment, Vol. \vldbvolume, No. \vldbissue\ %
ISSN 2150-8097. \\
\href{https://doi.org/\vldbdoi}{doi:\vldbdoi} \\
}\addtocounter{footnote}{-1}\endgroup

\ifdefempty{\vldbavailabilityurl}{}{
\vspace{.2cm}
\begingroup\small\noindent\raggedright\textbf{PVLDB Artifact Availability:}\\
The source code, data, and/or other artifacts have been made available at \url{\vldbavailabilityurl}.
\endgroup
}

\begin{figure}[t]
    \centering
    \begin{subfigure}[b]{0.39\linewidth}
        \centering
        \includegraphics[width=\linewidth]{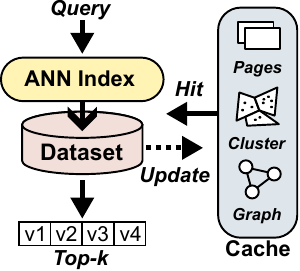}
        \caption[Traversal-centric]{Traversal-centric}
        \label{subfig:traversal-centric}
        \vspace{-1.5mm}
    \end{subfigure}
    \hspace{0.08\linewidth}
    \begin{subfigure}[b]{0.485\linewidth}
        \centering
        \includegraphics[width=\linewidth]{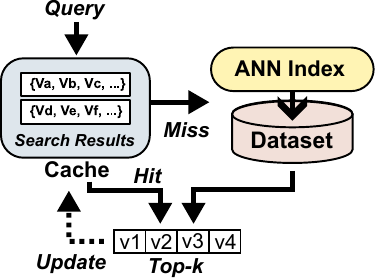}
        \caption[Result-centric]{Result-centric}
        \label{subfig:result-centric}
        \vspace{-1.5mm}
    \end{subfigure}
    \caption{
        \textbf{Two categories of caching strategies for ANNS.}
    }
    \label{fig:cache-cat}
    \vspace{-3mm}
\end{figure}

\section{Introduction} \label{sec:introduction}

Vector search has become a foundational component in modern data-intensive applications, spanning web search engines~\cite{chen2021spann, spacev}, recommendation systems~\cite{koenigstein2012efficient, li2017fexipro, lian2020lightrec, xiao2022progressively, covington2016deep}, scientific discovery~\cite{bajusz2015tanimoto}, and retrieval-augmented generation (RAG)~\cite{achiam2023gpt, lewis2020retrieval, bergman2025leveragingmid}.
These workloads rely on approximate nearest neighbor search (ANNS) to efficiently probe million- to billion-scale vector collections and retrieve relevant results under sub-second latency constraints.
At the core of ANNS is the \textit{vector index}, which organizes high-dimensional vectors according to their proximity, enabling efficient query processing while preserving high search accuracy.

As data volumes continue to grow, modern vector stores place both vectors and indexes on disk at low cost~\cite{pgvector, wang2024starling, chen2024singlestore, azurepgvector, awspgvector, xu2023spfresh, jayaram2019diskann, chen2021spann}.
However, disk-resident storage makes I/O the dominant performance bottleneck during vector search.
We observe that this bottleneck primarily stems from the inherent \textit{read amplification} in ANNS: to satisfy accuracy requirements, a query typically accesses orders of magnitude more data than is ultimately returned.
Specifically, the query first traverses a disk-resident index -- organized as clusters~\cite{douze2024faiss, chen2021spann, xu2023spfresh} or graphs~\cite{malkov2018efficient, jayaram2019diskann} -- during which candidates are fetched from disk into memory for distance computation, and then re-ranks the visited candidates to produce the top-$k$.
Achieving higher accuracy targets 
requires more aggressively expanding the search scope during index traversal, which substantially increases disk I/O and further exacerbates read amplification~\cite{yin2025panns}.

To mitigate this bottleneck, prior work has improved ANNS performance through algorithmic optimizations~\cite{chen2021spann, jayaram2019diskann, xu2023spfresh, yin2025panns} or emerging hardware~\cite{tian2024scalable, ren2020hm, tian2025towards, jang2023cxl}. 
Complementary to these efforts, this paper investigates a practical result-centric cache for disk-based ANNS, with the goal of reusing past query results in DRAM to improve end-to-end query performance.

Figure~\ref{fig:cache-cat} summarizes two caching strategies for ANNS: \textit{traversal-centric} caching and \textit{result-centric} caching.
Traversal-centric caches~\cite{ren2020hm, jayaram2019diskann, pgvector, yang2020pase} retain intermediate data accessed during index traversal in memory, such as pages, clusters, or graph neighbors.
By reusing the cached index structures and vectors, this approach reduces disk I/O during query processing.
However, sustaining high performance typically requires a large cache budget, as ANNS traversal often incurs a high memory footprint due to excessive vector accesses.

In contrast, result-centric caches~\cite{pandey2009nearest,falchi2008metric,chierichetti2009similarity,salem2022ascent,selvam2025simcache} store only the past query vectors and their top-$k$ results.
To serve an incoming query, the system evaluates its similarity to cached queries and returns the corresponding top-$k$ results when the similarity exceeds a predefined threshold.
As this strategy does not require strict query equivalence, it constitutes an \textit{approximate cache} relying on semantic similarity between queries.
By avoiding the storage of intermediate traversal data, result-centric caching significantly reduces memory consumption.
Moreover, cache hits can entirely bypass expensive ANN index traversals, yielding substantial performance gains.
As a result, result-centric caching represents a promising direction for improving ANNS efficiency.
However, the broader hit predicate makes cache hits no longer accuracy-safe, 
introducing two challenges that limit existing designs in practice.

\smallskip
\noindent
\textbf{Dilemma between cache efficiency and accuracy.}
Since result-centric caching relies on query similarity, it faces a trade-off between cache efficiency and accuracy.
A loose similarity threshold increases cache hit ratios and thus improves system efficiency, but may match insufficiently similar queries, degrading accuracy.
Conversely, a tight threshold preserves accuracy but may lead to low cache hit ratios by missing potential reuse opportunities.
Furthermore, vector distributions in the high-dimensional space are often highly skewed: some regions are dense, while others are sparse.
This heterogeneity renders the one-size-fits-all similarity thresholds adopted by prior work~\cite{bergman2025leveragingmid, guo2018potluck, selvam2025simcache, pandey2009nearest, falchi2008metric, chierichetti2009similarity} ineffective in balancing efficiency and accuracy across varying local vector densities.

\smallskip
\noindent
\textbf{Heavyweight cache refresh.}
Modern vector databases continuously receive update operations~\cite{xu2023spfresh, singh2021freshdiskann}, requiring cached results to be refreshed to avoid staleness and maintain correctness.
Otherwise, caches may miss newly inserted vectors or return deleted items.
However, supporting cache refresh is expensive: it either requires scanning all cached entries to determine whether their results should be updated or aggressively invalidating cached entries, which forces costly ANNS executions in the underlying database.
Consequently, existing result-centric caching approaches~\cite{guo2018potluck, bergman2025leveragingmid} choose to permit stale entries at the expense of ANN accuracy or degrade cache efficiency under frequent updates.

\smallskip
This paper presents \aker, an approximate cache that makes similarity-based reuse practical 
for disk-based vector searches by balancing accuracy and efficiency while supporting lightweight cache refresh.
\aker~is built on two core ideas.
First, \aker~introduces a \textit{density-aware caching strategy} which maintains an individual similarity threshold for each cached query.
Specifically, \aker~adapts the threshold based on the local data distribution: it tightens the threshold for queries located in dense regions to preserve accuracy, while relaxing it in sparse regions to improve cache utilization.
Moreover, the similarity threshold is dynamically adjusted at runtime, with accuracy treated as a first-class optimization objective.
Second, \aker~introduces \textit{del-consistency}, a practical consistency model between the cache and the underlying vector database.
Del-consistency enforces immediate invalidation of deleted items to prevent stale results from being returned, while allowing the visibility of insertions to be deferred for update efficiency, 
which aligns well with the approximate nature of ANNS.

We implement \aker~through carefully designed caching structures that achieve high memory efficiency and fast cache-side searches.
To reduce metadata overhead for queries sharing identical top-$k$ results, \aker~organizes cached queries into representative entries and lightweight alias entries.
Only representative entries store the top-$k$ results, while alias entries are linked to their corresponding representatives.
To support efficient lookup for similar queries, \aker~maintains a query filter which builds an embedded ANN index over cached queries, enabling the retrieval of similar queries without brute-force scans.
Finally, \aker~implements del-consistency via a lightweight refresh pipeline.
We execute deletions in the cache by removing deleted vectors from cached results and repairing affected top-$k$ lists using a small pool of top-$(k + \Delta)$ candidates, avoiding costly ANNS traversal in the underlying database.
In contrast, insertion visibility is safely deferred by tracking an accuracy degradation risk for each cached query, triggering top-$k$ recomputation only when the estimated risk exceeds a threshold.

We make three contributions.
First, we identify two fundamental limitations of existing approximate caches for ANNS and highlight the design challenges.
Second, we propose \aker, an approximate caching system that combines density-aware similarity thresholds to balance accuracy and efficiency with del-consistency to efficiently support cache refresh under continuous updates.
Finally, we implement \aker~with memory-efficient caching structures that support fast lookups and lightweight refresh pipelines.

By integrating \aker~into the popular PostgreSQL pgvector extension~\cite{pgvector, awspgvector, azurepgvector, gcppgvector}, we verify the effectiveness of \aker~ through end-to-end evaluations.
Across both web search and RAG workloads, \aker~improves accuracy by up to $64$ percentage points over prior state-of-the-art approximate caches.
Moreover, compared to pgvector’s default shared buffer (i.e., traversal-centric caching), \aker~reduces read amplification by up to $0.4\times$ and increases throughput by up to $3.2\times$ while using only $0.6\times$ of the memory.

\section{Background} \label{sec:background}

\subsection{Vector Search and Index} \label{subsec:vector-search}

Vector search is widely used in modern data-intensive applications such as web search~\cite{chen2021spann, spacev} and retrieval-augmented generation~\cite{achiam2023gpt, lewis2020retrieval, bergman2025leveragingmid}.
In these systems, each data item is represented as a vector in a high-dimensional space.
Given a query vector $q$, the system retrieves the top-$k$ most similar vectors according to a similarity metric\footnote{We use similarity and distance interchangeably in this paper.
Higher similarity between vectors indicates smaller distance in the vector space.} such as Euclidean, cosine, or inner product~\cite{chen2021spann, bergman2025leveragingmid, pgvector, douze2024faiss}.

Finding nearest neighbors via a brute-force scan yields exact results but incurs prohibitively high computational cost, especially at large scale.
Consequently, modern vector stores adopt approximate nearest neighbor search (ANNS) to trade result accuracy for high search efficiency~\cite{jayaram2019diskann, chen2021spann, pgvector, douze2024faiss, xu2023spfresh, bergman2025leveragingmid}.
ANNS systems rely on vector indexes that organize vectors based on proximity for efficient retrieval of target vectors.
Existing ANN indexes can be broadly categorized into two classes: cluster-based~\cite{douze2024faiss, pgvector, chen2021spann, xu2023spfresh} and graph-based~\cite{jayaram2019diskann, singh2021freshdiskann, malkov2018efficient, ren2020hm, wang2024starling, tian2024scalable}.
Cluster-based indexes group similar vectors using clustering algorithms such as $k$-means~\cite{douze2024faiss, chen2021spann, xu2023spfresh}, with each cluster represented by a centroid.
At query time, the system probes a small subset of clusters whose centroids are closest to the query vector.
In contrast, graph-based indexes (e.g., HNSW~\cite{malkov2018efficient}) construct a proximity graph in which each vector is connected to its nearest neighbors.
Queries are processed via greedy graph traversal starting from one or more entry points, progressively moving toward closer vectors while pruning unrelated regions of the graph.

The accuracy of ANNS is commonly evaluated using 
$recall@k = \frac{|R_a \cap R_e|}{k}$, 
where $R_a$ denotes the set of $k$ searched vectors and $R_e$ denotes the ground-truth top-$k$ results.
In practice, ANN indexes expose a trade-off between search accuracy and efficiency metrics such as throughput (QPS) and latency~\cite{douze2024faiss, malkov2018efficient, chen2021spann, jayaram2019diskann, xu2023spfresh, yin2025panns}.
Owing to their cost-effectiveness and competitive performance, disk-based ANNS solutions are widely adopted in vector databases~\cite{wang2021milvus, pgvector, weaviate, redis, qdrant} and have been deployed in major cloud services~\cite{azurepgvector, awspgvector}.

\vspace{-0.3em}
\subsection{Caching Strategy for Vector Search} \label{subsec:data-and-res-caching}

To reduce I/O costs, disk-based ANNS systems employ DRAM to cache frequently accessed data~\cite{xu2023spfresh, ren2020hm, chen2021spann, jayaram2019diskann, wang2024starling, pgvector}.
As illustrated in Figure~\ref{fig:cache-cat}, existing caching mechanisms for ANNS can be broadly categorized into two types -- \textit{traversal-centric} caching and \textit{result-centric} caching -- based on the type of data being cached.

Traversal-centric caches store index structures and vectors accessed during ANNS traversal.
For instance, in cluster-based ANN indexes, the cache typically contains centroids and accessed clusters, whereas in graph-based indexes it stores frequently visited neighbors.
Some approaches adopt static caching strategies~\cite{jayaram2019diskann}, which place vectors near the entry points directly in DRAM, without cache replacement between disk and memory.
Other systems, such as PostgreSQL pgvector~\cite{pgvector}, leverage the traditional database buffer pool to dynamically cache the hot data.
Traversal-centric caching is effective when the dataset or index structures are small enough to fit in memory, or when sufficient DRAM is available to cache a substantial fraction of the traversal-related data.

In contrast, result-centric caches aim to reduce memory consumption by storing only previously issued query vectors along with their final top-$k$ results.
When processing an incoming query, the system computes its similarity to cached queries and directly returns the corresponding top-$k$ results if the queries match exactly or if their similarity meets a predefined threshold.
Since the value of $k$ is typically fixed by the application and queries are often skewed, result-centric caching has proven effective in practice and been adopted in various applications~\cite{cambazoglu2012cache,guo2018potluck,finamore2022accelerating}.

\vspace{-0.3em}
\subsection{Approximate Cache} \label{subsec:approx-caches}
Beyond serving equal-match queries, result-centric caches for ANNS provide a unique opportunity to reuse search results across queries that are semantically similar~\cite{teevan2007information}.
This opportunity arises from a fundamental property of modern embedding models: vectors that are close in the embedding space typically correspond to semantically similar texts or images.
Consequently, even when two query vectors are not identical but sufficiently close, their top-$k$ nearest neighbors in the database often exhibit high overlap.

This observation enables result-centric caches to operate as an \textit{approximate cache}, in which the result returned from the cache may differ from that produced by the underlying ANNS engine.
In such a cache, query similarity serves as the cache-hit predicate.
Let $q$ denote a cached query and $A$ its associated top-$k$ results.
When a new query $q'$ arrives, the cache returns $A$ if the distance between $q$ and $q'$, denoted by $d(q, q')$, falls below a predefined threshold $\tau$.
The choice of $\tau$ is therefore critical, as it directly governs the trade-off between result accuracy and cache efficiency (hit ratios).

We present two representative approximate cache designs: \textsc{Proximity}~\cite{bergman2025leveragingmid} and Potluck~\cite{guo2018potluck}.
\textsc{Proximity}~\cite{bergman2025leveragingmid} is an approximate cache designed for RAG systems, where each cache entry stores a query vector along with its top-$k$ search results.
Cache hits are determined using a fixed similarity threshold $\tau$, which is configured at system initialization and remains unchanged during runtime.
While this design is simple and easy to deploy, the static threshold makes it difficult to balance accuracy and efficiency under varying query workloads and data distributions.

Potluck~\cite{guo2018potluck} is an approximate cache originally proposed to share computation results across mobile applications, but it is general enough to apply to ANNS workloads.
Unlike \textsc{Proximity}, Potluck dynamically adjusts the similarity threshold during runtime to improve cache effectiveness.
To this end, Potluck introduces a \textit{dropout} mechanism to estimate the accuracy of cached results.
When a query hits the cache, Potluck probabilistically triggers an intentional cache miss and executes the query against the underlying ANNS engine.
The returned result is then compared with the cached result.
If the two results match, indicating high cache accuracy, $\tau$ is increased to improve the cache hit ratio for future queries; otherwise, it is decreased to reduce approximation error.
\section{Motivation and Challenges} \label{sec:motiv}

\subsection{Read Amplification in ANNS} \label{subsec:read-amp}

We observe that the I/O bottleneck in disk-based ANNS primarily stems from excessive vector accesses during index traversal, particularly when high recall is required.
To quantify this effect, we define read amplification $\rho$ as the ratio between the total size of vectors accessed during index traversal, denoted by $C$, and the size of the top-$k$ results read from disk, denoted by $R$:
$\rho \triangleq \frac{C}{R}$.
For traditional indexes over scalar data types, such as LSM-trees, read amplification is effectively bounded on the order of a few tens (e.g., $20$–$35\times$) due to simplified traversal paths~\cite{luo2023smart, zhang2022depart, sarkar2021constructing, cao2020characterizing, olzhas2019slm}.
In contrast, disk-resident ANNS exhibits substantially higher read amplification, as vector queries often need to access a large number of neighboring vectors to avoid missing potential candidates.

To study this behavior, we measure the read amplification of two representative disk-based ANNS systems -- PostgreSQL pgvector~\cite{pgvector} and SPANN~\cite{chen2021spann} -- using the BIGANN~\cite{bigann} and SPACEV~\cite{spacev} datasets.
We cover two vector indexes supported in pgvector: the graph-based HNSW and the cluster-based IVF.
During query processing, pgvector relies on PostgreSQL’s shared buffer pool to load index structures and vector data from disk and cache frequently accessed pages.
Therefore, the shared buffer pool effectively functions as a traversal-centric cache.
SPANN~\cite{chen2021spann} is a cluster-based system that adopts a static caching strategy, where cluster centroids are retained in memory while all posting lists reside on disk.

\begin{figure}[t]
    \centering
    \begin{subfigure}[b]{0.495\linewidth}
        \centering
        \includegraphics[width=\linewidth]{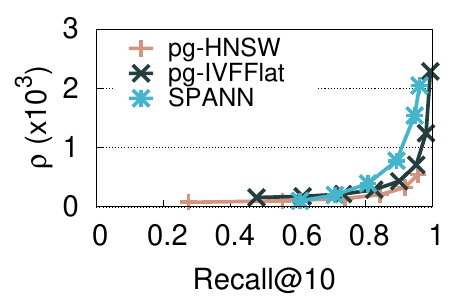}
        \caption[Dataset: BIGANN]{Dataset: BIGANN~\cite{bigann}}
        \vspace{-1mm}
        \label{subfig:ra-bigann}
    \end{subfigure}
    \hfill
    \begin{subfigure}[b]{0.46 \linewidth}
        \centering
        \includegraphics[width=\linewidth]{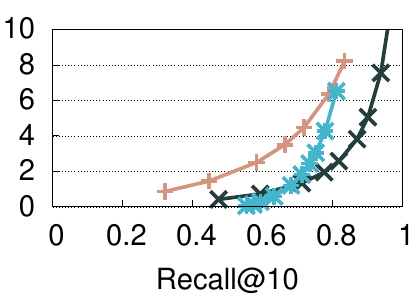}
        \caption[Dataset: SPACEV]{Dataset: SPACEV~\cite{spacev}}
        \vspace{-1mm}
        \label{subfig:ra-spacev}
    \end{subfigure}
    \caption{
    Read amplification ($\rho$) of disk-based ANNS systems under different recalls. The amplification becomes extremely severe at high recall requirements 
    (e.g., $0.8$-$0.9$).
    }
    \label{fig:ra}
    \vspace{-2mm}
\end{figure}

\begin{figure}[t]
    \centering
    \begin{subfigure}[b]{0.525\linewidth}
        \centering
        \includegraphics[width=\linewidth]{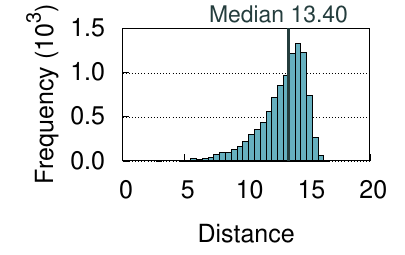}
        \caption[Dataset: BIGANN]{Dataset: BIGANN}
        \vspace{-1mm}
        \label{fig:bigann-gt-margin}
    \end{subfigure}
    \begin{subfigure}[b]{0.455 \linewidth}
        \centering
        \includegraphics[width=\linewidth]{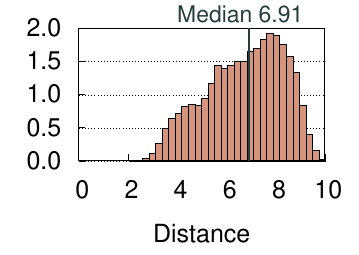}
        \caption[Dataset: SPACEV]{Dataset: SPACEV}
        \vspace{-1mm}
        \label{fig:spacev-gt-margin}
    \end{subfigure}
    \caption{
        Distribution of ground-truth top-1 distances to the queries.
        Large distance implies even the nearest neighbor is far away; thus the query lies in a sparser region.
    }
    \label{fig:gt-gap}
    \vspace{-2.5mm}
\end{figure}

Figure~\ref{fig:ra} reports the read amplification of different indexes under varying recall requirements.
Even at a low recall of $0.5$, pgvector’s HNSW (pg-HNSW) and IVFFlat (pg-IVFFlat) already exhibit significant read amplification, reaching $70\times$ and $155\times$ across both datasets.
Read amplification worsens as the recall requirement increases.
To improve recall from $0.5$ to $0.8$, pg-HNSW incurs at least $7.4\times$ and $9.5\times$ higher disk I/Os on the two datasets, while pg-IVFFlat requires $14.7\times$ and $24.6\times$ more, respectively.
SPANN shows a comparable trend, suffering from similarly severe read amplification.
Therefore, meeting application-level recall targets (e.g., $0.8$–$0.9$) can result in read amplification on the order of $10^2$–$10^3\times$.

The root cause is that achieving high recall necessitates accessing a much larger number of candidate vectors to avoid missing true nearest neighbors.
For instance, on the SPACEV dataset, to return top-10 results occupying $4$\,KB, pgvector reads an average of $29.7$\,MB of data per query to achieve $recall=0.9$.
This mismatch highlights the inefficiency of disk accesses at high recall.
While in-memory ANNS solutions eliminate disk I/O by keeping all data in DRAM, they scale poorly to large datasets due to prohibitive memory costs~\cite{douze2024faiss,jayaram2019diskann,ren2020hm}.
As an alternative, caching query results in memory offers an attractive design point: upon a cache hit, the system can bypass index traversal entirely, thereby substantially reducing read amplification and end-to-end query latency.

\subsection{A Single Global Threshold for All Queries} 
\label{subsec:one-size-reuse}

Prior approximate caches rely on a single similarity threshold to determine whether an incoming query can become a cache hit.
The similarity threshold $\tau$ is either globally fixed, or periodically adjusted.
However, the single global $\tau$ struggles to simultaneously achieve high cache efficiency and accuracy.
This limitation stems from the inherent non-uniformity of vector spaces: the local vector density -- the number of vectors within a given radius $r$ -- can vary substantially across different regions~\cite{falchi2008metric}.
A global threshold implicitly assumes that a single reuse radius is suitable for all queries and enforces a trade-off.
Specifically, a larger $\tau$ improves cache hit ratios by treating more queries as hits, 
but risks returning incorrect neighbors in dense regions; conversely, a smaller $\tau$ preserves accuracy while causing unnecessary cache misses in sparse regions.

To quantify this effect, we empirically examine density variance using the BIGANN and SPACEV datasets.\footnote{
For simplicity and clarity of presentation, we use Euclidean distance throughout the paper, where smaller values indicate higher similarity between vectors.
Our design can be naturally extended to other distance metrics, such as cosine distance.
}
Because selecting a meaningful radius $r$ is non-trivial, we approximate local density using the distance from each query to its closest ground-truth neighbor.
For a query $q$, this distance is the radius of the smallest hypersphere centered at $q$ that contains at least one data point.
Accordingly, smaller distances indicate denser regions of the vector space, while larger distances correspond to sparser regions.

As shown in Figure~\ref{fig:gt-gap}, the top-1 distances range from $2.7$ to $16.7$ on BIGANN and from $4.2$ to $10.0$ on SPACEV, indicating substantial variation in local density.
A single threshold thus cannot adapt to diverse query characteristics.
In Section~\ref{subsec:eval-cache-acc}, we empirically verify that no single threshold can simultaneously balance recall and cache hit rate. 
This observation motivates our design of a \textit{density-aware similarity threshold}, which maintains per-query thresholds and hit conditions and dynamically adjusts each threshold at runtime.

\vspace{-0.2em}
\subsection{Absence of Cache Refresh Scheme}

When the underlying ANN indexes evolve due to insertions and deletions, the cache must reflect these changes to prevent staleness.
Existing approximate caching schemes typically assume a static dataset and therefore lack explicit cache refresh mechanisms.
However, real-world workloads may continuously process insert and delete operations to maintain dynamic vector collections~\cite{xu2023spfresh, li2018design, wei2020analyticdb}.
In those workloads, if a vector closer to a query $q$ is newly inserted, its cached top-$k$ results must be updated to preserve search quality.
Similarly, when one or more vectors in the cached top-$k$ are deleted, the cache must be refreshed to reflect the changes.

Refreshing cached results in approximate caches is non-trivial.
Consider a query $q$ whose cached top-3 results are ${v_a, v_b, v_c}$ with distances $0.13$, $0.24$, and $0.31$, respectively.
If a new vector $v_d$ is inserted with distance $0.19$ to $q$, the correct top-3 becomes ${v_a, v_d, v_b}$.
Updating the cache accordingly, however, can incur substantial overhead.
Specifically, the system must first determine which cached queries are affected by the insertion, which requires scanning all cached queries and computing their similarity to the newly inserted vector.
Deletions introduce analogous challenges: the system must identify cache entries whose top-$k$ results include deleted vectors.
Moreover, deletions invalidate cached queries because they reduce the number of available results below $k$.
This requires the system to re-execute ANNS in the underlying database to reconstruct the top-$k$.

To efficiently maintain cache consistency under dynamic updates, we propose a result-level consistency model, termed \textit{del-consistency}.
Based on this model, we further design a lightweight cache refresh mechanism that incurs low maintenance overhead while effectively bounding recall degradation.
\begin{figure*}[t]
    \centering
    \begin{subfigure}[b]{0.54\linewidth}
        \centering
        \includegraphics[width=\linewidth]{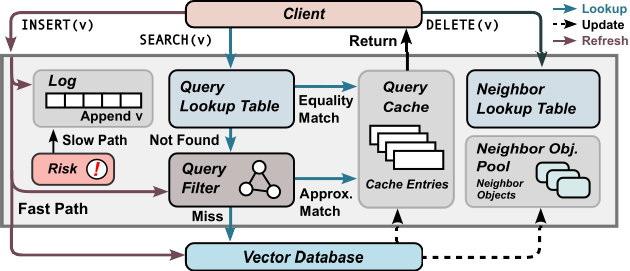}
        \caption[Architecture]{Architecture}
        \label{fig:arch-overview}
        \vspace{-2mm}
    \end{subfigure}
    \hspace{0.065\linewidth}
    \begin{subfigure}[b]{0.36\linewidth}
        \centering
        \includegraphics[width=\linewidth]{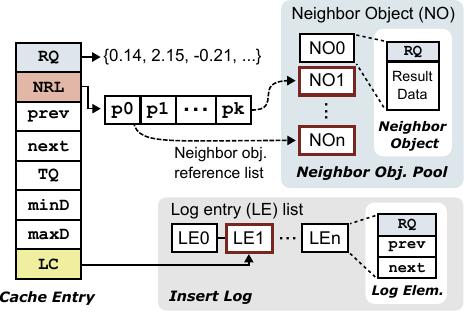}
        \caption[Cache entry structure]{Cache entry structure}
        \label{fig:entry-structure}
        \vspace{-2mm}
    \end{subfigure}
    \caption{\aker~overview.}
    \label{fig:overall-arch}
    \vspace{-6mm}
\end{figure*}

\section{\aker} \label{sec:topkache-design}
\subsection{Overview} \label{subsec:overview}

We present \aker, an approximate result cache for disk-based ANNS systems.
\aker~ combines density-aware reuse control with lightweight refresh to support similarity-based result caching under dynamic updates.
\aker~ provides the mechanisms required to make approximate caching practical inside a vector search engine.
Conventional caches are equivalence-based.
An exact key match is sufficient to locate cached data. 
Approximate caching enables broader reuse opportunities but introduces challenges that conventional caches do not encounter:
\textit{which cached query is similar enough to reuse}, 
\textit{how safe reuse is for that specific query},
and \textit{how cached results should remain meaningful under updates}. 
These requirements require additional components beyond a conventional cache.

First, \aker~ adds a \textit{query filter}, a small embedded ANN index over cached queries, with a low-cost management scheme 
because approximate reuse opportunities are hard to find through exact-key lookup alone.
Second, \aker~ adds fine-grained reuse control through an entry-specific threshold and its safety rule to adapt to the local neighborhood density.
Third, 
\aker~ adds a representative/alias cache entry structure and shared result management, 
because semantically similar queries may reuse the same top-$k$ result set 
and therefore require different metadata organization for space efficiency.
Finally, \aker~ adds \textit{del-consistency} implemented through risk-triggered fast and slow refresh mechanisms, 
because approximate cached results must remain useful under updates without forcing expensive re-traversal on the hit path.

Figure~\ref{fig:arch-overview} illustrates key components of \aker. 
For searches, 
\aker~ probes the cache and returns a cache entry on a hit; 
otherwise, it queries the vector database. 
For inserts and deletes, \aker~refreshes cached results to reflect updates in the base data.

\smallskip
\noindent
\textbf{Vector search.}
\aker~handles search requests with equality hits and approximate hits.
It first searches the query lookup table, a hash table that maps a hashed query vector to a cache entry.
Each cache entry stores the query vector and its top-$k$ results.
\aker~first searches for an identical query in the query lookup table.
When a match is found, the cache entry is returned.
Otherwise, \aker~takes an approximate search path by querying a query filter.

The query filter is a small ANN index built over the query vectors of the cache entries.
Its size depends on the number of cache entries.
The query filter serves two roles: 
(1) retrieving a cache entry whose query is similar to an incoming query without a brute-force scan, 
and (2) locating cache entries that are likely to be affected by update requests.
The cache entry is returned if an approximate hit condition is met; otherwise, the request becomes a miss.

\smallskip
\noindent
\textbf{Vector insertion.}
When a client inserts a new vector into the vector database, \aker~records the vector in an \textit{insert log}.
\aker~then refreshes affected cache entries via two paths: a \textit{fast path} and a \textit{slow path}.
The fast path uses the query filter to locate the cache entry whose query is closest to the inserted vector.
It checks whether the new vector would enter that cache entry's top-$k$ results; if so, \aker~updates the cached top-$k$ accordingly.
The slow path processes insert log records lazily in batches.
\aker~estimates the risk of recall degradation from lazy refreshes.
When the risk exceeds a predefined threshold, it refreshes unchecked log entries.
The two paths refresh the cache without a full scan of all cache entries.

\smallskip
\noindent
\textbf{Vector deletion.}
Upon a vector deletion request, \aker~ first searches a neighbor lookup table to check whether the vector is present in the cache.
Each neighbor object represents a cached result vector.
If any of the vectors requested to be deleted reside in the cache, \aker~immediately removes them and updates the neighbor lookup table.
Cache entries that fail to keep a complete set of top-$k$ are repaired during cache lookups by promoting reserved candidates.

\begin{table}[t]
\centering
\small
\caption{Summary of key designs in \aker.}
\label{tab:aker-terms}
\begin{tabular}{p{0.22\linewidth} p{0.57\linewidth} p{0.10\linewidth}}
\toprule
 & Description & Section \\
\midrule
Representative/ \newline alias cache entry
& Representative entries hold results; 
alias entries point to the representative entries.
& \ref{subsec:entry-str}, \ref{subsec:cache-lookup} \\
\midrule
Query filter
& Embedded index of cache entries that finds similar cache entry candidates.
& \ref{subsec:overview}, \ref{subsec:cache-lookup}, \ref{subsec:evict} \\
\midrule
Threshold $\tau_q$
& Reuse radius for a cached query $q$; enlarged and shrunk dynamically from hit types.
& \ref{subsec:entry-str}, \ref{subsec:past}, \ref{subsec:cache-lookup} \\
\midrule
Refresh
& Del-consistency combines immediate deletion with fast/slow refresh via risk tracking.
& \ref{subsec:overview}, \ref{subsec:refresh} \\
\midrule
Active/Standby \newline filter
& Two-filter scheme that separates cache entry insertions from evictions.
& \ref{subsec:evict} \\
\bottomrule
\end{tabular}
\end{table}

\subsection{Cache Entry Structure} \label{subsec:entry-str}

Figure~\ref{fig:entry-structure} shows the structure of a cache entry.
The raw-query field \texttt{RQ} stores the query vector.
The neighbor object reference list \texttt{NRL} stores references to the neighbor objects that store the query's top-$k$ results.
The \texttt{prev} and \texttt{next} pointers link similar cache entries.
The per-query similarity threshold $\tau_{q}$ is stored in the \texttt{TQ} field.
The \texttt{maxD} and \texttt{minD} fields keep the maximum and minimum distance between the query and its cached top-$k$, respectively.
\aker~initializes \texttt{TQ} to $\texttt{minD}/4$, a conservative setting that biases early lookups toward preserving recall until \aker~adapts \texttt{TQ} online.

Each element of \texttt{NRL} points to a neighbor object in a neighbor object pool.
\aker~allows multiple cache entries to share neighbor object references to save the cache space.
The cache size is controlled by capping the total number of neighbor objects.
The neighbor lookup table is a hash table that maps each key vector to its corresponding neighbor object in the shared pool.
Each neighbor object stores the raw vector and database-specific metadata (e.g., tuple identifiers in PostgreSQL).
Because a neighbor object can be referenced by multiple cache entries, \aker~ uses a reference counter for each neighbor object for safe memory reclamation.
When inserting a new cache entry, \aker~checks whether each top-$k$ vector already exists in the neighbor lookup table; if so, it reuses the existing neighbor object and increments its reference counter.
Otherwise, it creates a new neighbor object and inserts it into the table.
When a vector deletion is requested, \aker~removes the corresponding neighbor object from the table.

To avoid repeated approximate lookups for recurring queries, \aker~uses \textit{entry linking}.
Specifically, when a query is served via an approximate hit, \aker~registers a cache entry in the query lookup table.
This newly inserted entry points to the cache entry that was previously served.
When the same query vector arrives again, \aker~ directly returns the entry in $O(1)$ using the query lookup table, skipping both the query filter probe and the similarity check.

Cache entries are classified into two types in \aker: \textit{representative} and \textit{alias}.
A representative entry owns the \texttt{NRL} and is indexed in the query filter.
Each element in the representative entry \texttt{NRL} points to a neighbor object.
In contrast, an alias entry keeps an empty \texttt{NRL}.
It links to a representative entry and is not indexed in the query filter.
Cache entries are linked via \texttt{prev}/\texttt{next} pointers, forming a list whose head is the representative entry.
A cache lookup returns a representative entry either directly or by following \texttt{prev} from an alias entry to the list head.
By indexing only representative entries in the query filter, \aker~can keep the query filter small (Section~\ref{subsec:evict}), while allowing repeated queries that were previously served via approximate hits to be answered via the query lookup table.

The log checkpoint field \texttt{LC} references an insert log record.
It tracks the progress of slow path refresh for a cache entry, i.e., how far the cache entry has processed insert log updates.
When \aker~invokes the slow path after vector insertions, it performs batched distance checks, scanning from \texttt{LC} (Section~\ref{subsec:refresh}).
Only representative entries maintain a valid \texttt{LC}, and alias entries set \texttt{LC} to \texttt{null}.

\begin{figure}[t]
    \centering
    \includegraphics[width=0.94\linewidth]{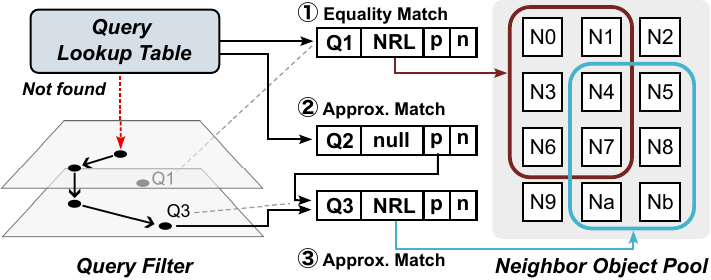}
    \caption[Search path]{
    \textbf{\aker~lookup}.
    \aker~probes query lookup table: on equality match (1) returns the entry; 
    on alias hit (2) follows \texttt{prev} to the representative. 
    Otherwise (3), probes query filter and returns a nearest cache entry; 
    otherwise it is a miss.}
    \label{fig:search-path}
    \vspace{-2mm}
\end{figure}

\vspace{-0.3em}
\subsection{Per-Query Similarity Threshold} \label{subsec:past}

\aker~maintains a per-query similarity threshold $\tau_q$ for each cached representative query $q$.
The threshold $\tau_q$ controls the acceptance region for approximate reuse around $q$.
When tuning $\tau_q$, \aker~ prioritizes both low-overhead control and conservative safeguards.
The most precise method is verification-based tuning.
A cache can re-traverse the index and adjust the threshold based on the search discrepancy ~\cite{guo2018potluck}.
However, in disk-based vector search systems, invoking the underlying ANNS on cache hits reduces the benefit of caching. 
Therefore, \aker~ avoids direct feedback that requires additional index traversals and instead uses a low-overhead signal already available in the cache: the hit type.

An equality hit implies that the incoming query exactly matches a cached query, 
whereas an approximate hit implies that the incoming query does not match any cached query exactly but is still close enough to a cached query to reuse its result. 
In \aker, the two hit types play different roles in threshold control.
An approximate hit is the primary warning signal that the current reuse radius may already be too permissive, 
because it serves one query using the cached result of another nearby query. 
\aker~ therefore shrinks $\tau_q$ after approximate hits. 
In contrast, an equality hit indicates that the current threshold has not introduced additional query-reuse error for that exact query.
\aker~ therefore treats equality hits as a low-risk signal for cautious increase in reuse opportunities.
Concretely, \aker~ updates the per-query threshold \texttt{TQ} as follows:
\begingroup
\begin{equation}
    \tau_{q, \textnormal{next}} \leftarrow 
        \begin{cases}
            \tau_{q, \textnormal{prev}} \cdot \alpha_L , & ~( \alpha_L > 1),~\textnormal{Equality-match}\\
            \tau_{q,\textnormal{prev}} \cdot \alpha_T , & ~(0 < \alpha_T <1),~\textnormal{Approximate-match}
        \end{cases}
\end{equation}
\endgroup
\noindent
Here, $\tau_q$ denotes the similarity threshold associated with a cached query $q$.
$\alpha_L > 1$ and $\alpha_T \in (0,1)$ are multiplicative factors for increasing and decreasing the threshold, respectively.
To achieve better hit ratios, $\tau_q$ must be large.
To achieve better search accuracy, $\tau_q$ must be small.
\aker~ prioritizes search accuracy, thus using asymmetric update factors $\alpha_L=1.1$ and $\alpha_T=0.25$ as its default values.
To prevent $\tau_q$ from growing far beyond a reasonable distance, \aker~limits the maximum $\tau_q$ by the distance between $q$ and its closest cached neighbor.
\aker~ permits only mild growth after equality hits, 
and this growth is capped by the distance to the closest cached neighbor. 
Repeated equality hits thus do not drive unbounded threshold expansion.
Each $\tau_q$ is updated on every cache hit.

\subsection{Cache Lookup} \label{subsec:cache-lookup}

Figure~\ref{fig:search-path} and Algorithm~\ref{alg:cache-entry-get-version2} present a cache lookup workflow. 
\aker~ distinguishes two cache-hit types: \textit{equality match} and \textit{approximate match}.
An equality match happens when an incoming query is identical to the cached query vector.
In the case of an equality match, \aker~first probes the query lookup table (line~\ref{algo-1:query_lookup_table}).
If an identical query vector is found, the corresponding cache entry is returned.

An approximate match occurs when the incoming query is not identical to the cached representative query but is sufficiently close.
\aker~supports two approximate match paths: (i) returning an alias entry via the query lookup table, or (ii) returning a representative entry via the query filter.
First, the alias path handles queries that have previously produced an approximate hit.
Suppose an incoming query $q'$ approximately matches a cached query $q$ stored in a representative entry $e$.
\aker~ creates an alias entry $e'$ for $q'$ and registers it in the query lookup table.
The alias entry $e'$ stores a reverse pointer \texttt{prev} to its representative entry $e$.
Upon a repeated request for the same vector $q'$, \aker~finds $e'$ from the query lookup table and follows \texttt{prev} to retrieve $e$ (line~\ref{algo-1:linked-exact-match}).
This avoids rerunning the query filter search and distance comparison because $q'$ is already similar to the previously validated representative entry.

\LinesNumbered
\RestyleAlgo{algoruled}
\normalem
\setlength{\textfloatsep}{0.35em}
\begin{algorithm}[t]
\DontPrintSemicolon
\caption{Cache Lookup}
\label{alg:cache-entry-get-version2}
\SetAlgoNoLine
\SetKwFunction{FnWriteReq}{cacheRead}
\SetKwComment{Comment}{//~}{ }
{
    \KwIn{query $q$}
    \SetKwProg{Fn}{Procedure}{:}{end}
    \Fn{\FnWriteReq{q}} {
        \( e \gets \textnormal{QueryLookupTable}(q) \)\; \label{algo-1:query_lookup_table}
        \If{\( e \) == \texttt{NULL}} {
            \( e \gets \textnormal{QueryFilter}(q) \) \; \label{algo-1:sim-match}
            \If{\( \texttt{distance}(q, e.RQ) < e.TQ \)}{
                \( \ e.TQ \gets \texttt{updateTQ}(e.TQ) \) \;
                $\Return ~e$ \;
            }
            $\Return ~\texttt{NULL}$ \olive{\Comment{miss}}
        }
        \While {\( e.prev \) is not \texttt{NULL}} {
            \( e \gets e.prev \) \olive{\Comment{Find representative entry}}
            \label{algo-1:linked-exact-match}
        }

        \( \ e.TQ \gets \texttt{updateTQ}(e.TQ) \) \;
        $\Return ~e$
    }
}
\end{algorithm}

If no alias entry is found in the query lookup table, \aker~searches the query filter for a similar representative entry (line~\ref{algo-1:sim-match}).
Only representative entries are indexed in the query filter.
Given a query $q'$, the query filter returns the nearest representative entry $e$ whose cached query vector \texttt{RQ} is closest to $q'$.
\aker~then computes the distance between $q'$ and \texttt{RQ}.
If the distance is below the per-query similarity threshold \texttt{TQ}, it becomes an approximate hit.

\subsection{Cache Update} \label{subsec:cac-writes}

To insert a new cache entry, 
\aker~ first generates neighbor objects for the retrieved top-$k$ results and registers them in the neighbor lookup table.
If an identical vector exists, \aker~increments the reference counter of the existing neighbor object.
\aker~then creates a new representative entry and records the references in its \texttt{NRL}.
The representative entry is registered in the query lookup table and indexed by the query filter.
If the request is an approximate match, \aker~creates an alias entry.
\aker~traverses the representative entry's \texttt{next} pointers and links the alias entry at the tail.
Finally, \aker~inserts the alias entry into the query lookup table.

\aker~initializes the \texttt{LC} to the insert log tail.
This is safe because the new representative entry is built over the latest database snapshot.
In the slow refresh path, \aker~scans the insert log starting from the cache entry's \texttt{LC} to determine whether the vector should be included in the top-$k$ set.
A representative entry created on a cache miss at time $t$ is computed over the database snapshot at time $t$.
Thus, the search result reflects all latest inserts up to the insert log tail at time $t$.

\vspace{-0.3em}
\subsection{Cache Refresh} \label{subsec:refresh}

\noindent
\textbf{Del-consistency.}
We define \textit{consistency} for ANNS caches as the result-level visibility agreement between the vectors in the underlying database and the vectors in the cache at time \(t\) for a query \(q\).
Let $V_t$ denote the set of live vectors in the database at time $t$, and let $c_t$ denote the cached top-$k$ results of $q$ returned on a cache hit.
Let $s_t$ denote the results of $q$ returned by executing the base ANNS on the database at time $t$.
We propose a new consistency model called \textit{del-consistency}.
A cache is \textit{del-consistent} at time $t$ for any query $q$
if it satisfies validity, while allowing relaxed visibility:
\theoremstyle{definition}
\newtheorem{prop}{P}
\begin{prop}[\textsc{Validity}]
At any time $t$, cached results of $q$ contain only valid vectors, i.e., $c_t \subseteq V_t$ and $|c_t| = k$.
\label{prop:corr}
\end{prop}
\begin{prop}[\textsc{Relaxed Visibility}]
At any time $t$, the cached results $c_t$ 
are not required to equal the ANNS results $s_t$ for the same query.
\label{prop:relv}
\end{prop}

\textit{Validity} requires that deletion operations be reflected immediately in \texttt{NRL}.
Otherwise, the cache could return vectors that no longer exist in the database.
\textit{Relaxed visibility} allows temporary staleness with respect to newly inserted vectors, 
which aligns well with ANNS as it returns approximate results.
\aker~ enforces del-consistency that guarantees validity under deletions without requiring eager recomputation of the latest top-$k$, which would impose high refresh cost on the hit path.

Consider a cache that follows the del-consistency model.
A cache entry for query $q$ is created at time $t_0$. 
If a cached neighbor $v$ is deleted at $t_1$ ($t_1 > t_0$), 
the cache removes $v$ immediately (P\ref{prop:corr}). 
If a new vector $u$ is inserted at $t_2$ ($t_2 > t_0$) and would appear in $s_{t_2}$, 
$u$ may remain invisible to $q$'s cache entry until the next refresh at $t_3$ ($t_3>t_2$).
During this interval, the cache can still return a valid top-$k$ (P\ref{prop:relv}).
\aker~ implements del-consistency by handling deletions immediately and insertions lazily, with a reserve parameter $\Delta$.

\smallskip
\noindent
\textbf{Deletions.}
\aker~immediately removes deleted vectors from the neighbor object pool to satisfy validity (P\ref{prop:corr}).
However, after removals, a cache entry may no longer contain $k$ valid neighbors.
In this case, the entry (cached query) must be invalidated.
In \aker, a neighbor object can be referenced by multiple cache entries, thus a single vector deletion may invalidate multiple entries.

To reduce cache entry invalidations, \aker~stores the top-$k$ results with an additional $\Delta$-sized reserve in \texttt{NRL}.
We choose $\Delta$ to be small and collect the reserve using the same search path as for the top-$k$.
ANNS algorithms typically maintain an internal candidate set that is larger than $k$ during search, 
even though they return only the top-$k$ results~\cite{malkov2018efficient, chen2021spann, macdonald2021approximate, bergman2025leveragingmid}.
Thus, \aker~ can save the $(k+\Delta)$ neighbors from the same traversal with negligible additional overhead.
When a deletion invalidates a neighbor object, \aker~removes it immediately and promotes the next closest reserved neighbor in \texttt{NRL} (P\ref{prop:corr}, P\ref{prop:relv}).
If deletions break \texttt{NRL} such that the number of valid neighbor objects falls below $k$, 
\aker~invalidates the corresponding cache entry.

\LinesNumbered
\RestyleAlgo{algoruled}
\normalem
\setlength{\textfloatsep}{0.35em}
\begin{algorithm}[t]
\DontPrintSemicolon
\caption{Cache Refresh (Insertions)}
\label{alg:wlog-insert}
\SetAlgoNoLine
\SetKwFunction{Insert}{fastPath}
\SetKwFunction{Batch}{slowPath}
\SetKwComment{Comment}{//~}{ }
    \KwIn{new vector $v$}
        Log $\mathcal L$;
        Batch size $B$;
        Risk $R$;
        ~~ Risk threshold $\tau_{\text{risk}}$; \;
    \SetKwProg{Fn}{Procedure}{:}{end}
    \Fn{\Insert{$v$}}{
        $e \gets$ \textnormal{QueryFilter}$(v)$ \olive{\Comment{fast path}}
        \label{algo-2:approx-get}
        \If{$\texttt{distance}(e,v) < e.\textit{maxD}$} 
        {
            $e.NRL \gets \texttt{updateNRL}(v)$\; 
            \label{algo-2:max-check}
        }
        $R \gets \texttt{updateRisk}(e)$\;  
        \If{$R$ $\ge\tau_{\text{risk}}$}{
              \Batch{$B$} \label{algo-2:slow-path}
        }
    }
    \smallskip
    \SetKwProg{Pn}{Procedure}{:}{end}
    \Pn{\Batch{$B$}}{
        $e_c \gets N extEntry$ \olive{\Comment{e\_c: Cache entry}} \label{algo-2:rr}
        \For{$e_l = \mathcal L$ $[e_c.LC]$ \KwTo $\mathcal L$ $[e_c.LC + B]$}{ 
            \If{$\texttt{distance}(e_l.RQ, e_c.RQ) < e_c.\textit{maxD}$}
            { \label{algo-2:slow-check}
                   $e_c.NRL \gets \texttt{updateNRL}(e_l)$ \olive{\Comment{e\_l: Log entry}}
            }
        }       
        $e_c.LC \gets e_{c}.LC + B$\;
        $R \gets \texttt{updateRisk}(e_c)$\;   
    }
\end{algorithm}

\smallskip
\noindent
\textbf{Insertions.}
When a new vector is inserted into the database, it is unknown which cache entries should include it in their \texttt{NRL}.
A naive approach that scans every cached query and updates affected \texttt{NRL}s is expensive.
To minimize overhead, \aker~splits insertion handling into a \textit{fast path} and a \textit{slow path} (Algorithm~\ref{alg:wlog-insert}).

The fast path targets the cache entry that is most likely to be affected by the insertion (P\ref{prop:relv}).
Upon receiving an insert request, \aker~appends the new vector to the insert log and queries the query filter (line~\ref{algo-2:approx-get}) to retrieve the closest representative entry.
\aker~then compares the distance between the inserted vector and the entry's \texttt{RQ} against the entry's \texttt{maxD} (line~\ref{algo-2:max-check}).
If the inserted vector is closer than \texttt{maxD}, \aker~evicts the farthest neighbor from \texttt{NRL}, inserts a reference to the new vector, and updates \texttt{maxD} accordingly.

Because the fast path updates only one entry per insertion, other entries that should also include the new vector may remain stale.
To propagate insertions beyond the fast path target, the slow path incrementally refreshes the remaining entries (line~\ref{algo-2:slow-path}).
On each cache lookup or insert log append, \aker~selects one representative entry in a round-robin manner (line~\ref{algo-2:rr}).
For the selected entry, \aker~scans the next batch of log records (default $16$) starting from its \texttt{LC}.
For each scan, \aker~computes the distance between the inserted vector and \texttt{RQ} (line~\ref{algo-2:slow-check}) and substitutes the inserted vector in the same manner as the fast path.
Then, \aker~advances \texttt{LC} to the most recently examined log position and recomputes the global \textit{risk} $\textnormal{R}$.

During vector insertions and cache lookups, 
\aker~triggers the slow path when $\textnormal{R}$ exceeds a predefined risk threshold.
\aker~maintains $\textnormal{R}$ as a conservative estimate of potential recall degradation under delayed insert propagation.
$\textnormal{R}$ is derived from per-entry \textit{risk factors} $r$ and the number of unchecked vectors in the insert log.
For each representative entry $e$, $r(e)$ estimates how likely an unseen inserted vector could enter $e$'s top-$k$ by replacing a member of \texttt{NRL}.
\aker~uses the $\Delta$ reserve region in \texttt{NRL} as a proxy for how tight the top-$k$ boundary is:
since only the top-$k$ results are returned, 
refreshing items at ranks $k{+}1$ to $k{+}\Delta$ can be delayed without changing the returned results (P\ref{prop:relv}).
\begingroup
\begin{equation}
    r(e) = \frac{d_k (e)}{d_{k+\Delta}(e)} \in [0,1],
    \quad
    \textnormal{R} =
    \overline{r} \cdot 
        \frac{\overline{U}}{W} 
\end{equation}
\endgroup
\noindent
Here, $d_{\alpha}(e)$ denotes the distance between $e$'s query and the $\alpha$-th neighbor in $e$'s cached top-$(k{+}\Delta)$ list.
A value of $r(e)$ close to $1$ indicates a higher likelihood that an unseen insertion could substitute into the top-$k$.
By design, \aker~over-estimates the risk by treating all unchecked insert log records as potentially harmful.
$\overline{r}$ is the average risk factor, 
$\overline{U}$ is the average number of unchecked insert log records per cache entry, 
and $W$ is the total insert log size.
$\textnormal{R}$ scales with both how tight the top-$k$ boundaries are and how far entries lag behind the insert log.
As $\textnormal{R}$ increases toward $1$, the potential for recall degradation increases.

\smallskip
\noindent
\textbf{Insert log.}
\aker~reclaims insert log space when evicting a cache entry 
whose \texttt{LC} references the log head.
Starting from the head, \aker~advances the head and removes log records that are no longer referenced by the \texttt{LC} of any cache entry.
This reclamation stops at the earliest log record that is referenced by at least one cache entry.

\subsection{Eviction} \label{subsec:evict}

\noindent
\textbf{Neighbor object pool.}
\aker~evicts neighbor objects when a new cache entry requires allocating new neighbor objects but the pool has no free space.
To free space, \aker~selects a victim representative entry and reclaims the neighbor objects referenced by its \texttt{NRL}.
We adopt FIFO because it has low overhead.
Prior work suggests that under tight similarity thresholds, the accuracy impact of the eviction policy is minor~\cite{bergman2025leveragingmid}.
Given a victim representative entry, \aker~traverses its \texttt{NRL} and decrements the reference count of each neighbor object.
If the reference count reaches zero, it is removed.
Otherwise, the object remains alive because it is still referenced by other cache entries.
If reclaiming a single victim entry does not provide enough free space, \aker~ evicts the next FIFO victim and continues until the space becomes sufficient.

When evicting a representative entry, \aker~also removes all linked alias entries.
It follows the representative's \texttt{next} pointers, removes each alias entry, and deletes the corresponding query lookup table records.
After all aliases are removed, \aker~deletes the victim representative entry and removes it from the query lookup table.
Finally, \aker~marks the representative entry as a tombstone in the query filter to prevent it from being returned by subsequent approximate lookups.

\begin{figure}[t]
    \centering
    \includegraphics[width=0.99\linewidth]{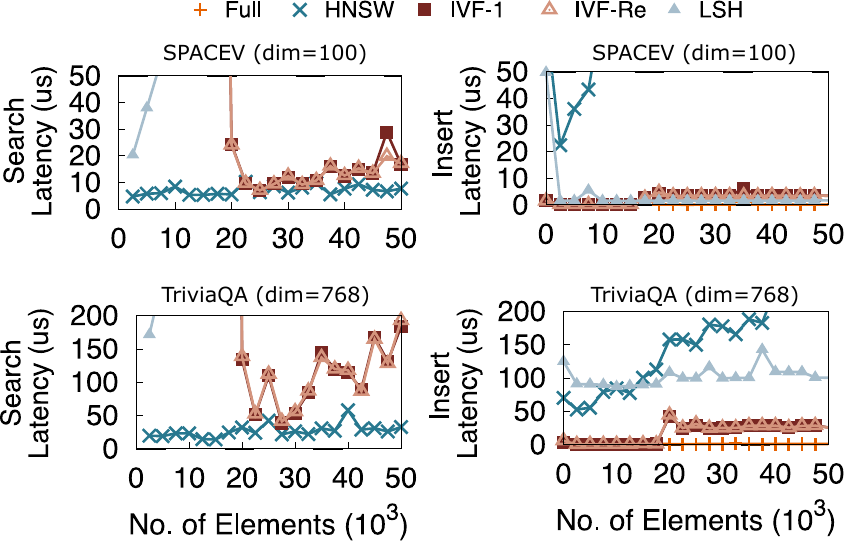}
    \caption[ANN operation latency]{
        ANN index operation latency.
            We grow indexes to measure latency of top-$1$ lookup and incremental builds.
            We compare FAISS~\cite{douze2024faiss} HNSW ($m=4$, $efConstruction=8$, $efSearch=16$), LSH ($nbits$ matched to vector dimensionality), 
            IVF with single build at 20K inserts (IVF-1; \texttt{IVFFlat} $nlists=128$, $nprobe=1$), and with periodic rebuild at every 10K inserts (IVF-Re), 
            and a brute-force scan (Full). 
    }
    \label{fig:rev-faiss-microbench}
    \vspace{-1.5mm}
\end{figure}

\smallskip
\noindent
\textbf{Query filter.}
The query filter is the embedded ANN index over representative cached queries.
To choose the best candidate for the query filter, we microbenchmark several index types~\cite{douze2024faiss} under an in-memory setting using two query sets (Section ~\ref{subsec:exp-setup}) as shown in Figure~\ref{fig:rev-faiss-microbench}.
The results show a clear trade-off between lookup and insertion latency across the index types.
HNSW provides the lowest lookup latency even at higher dimensionality, but its insertion latency grows with the number of indexed elements.
IVF shows low latency in insertion, but its lookup latency is higher than HNSW and grows more with the number of indexed elements.
LSH is slowest in lookup and is less aligned with the distance metric used by the underlying ANNS.
Because \aker~ targets read path optimization, we choose HNSW because it provides the lowest lookup latency.

The adoption of HNSW for the query filter introduces two major challenges.
First, to keep insertion overhead of the query filter low, \aker~ must bound the size of the index.
Second, graph-based ANN indexes like HNSW make fine-grained deletions non-trivial because the removal may break the graph connectivity~\cite{douze2024faiss, xu2023spfresh, jayaram2019diskann}.
Prior systems often implement deletions via tombstones and periodic rebuilds ~\cite{xu2023spfresh, jayaram2019diskann}.
This approach has three limitations.
First, searches may return deleted items before the next rebuild.
Second, tombstoned elements inflate the index size and increase insertion latency.
Finally, selecting the timing of index rebuilds becomes important, since rebuilding from scratch can amplify tail latency.

\begin{figure}[t]
    \centering
    \includegraphics[width=0.95\linewidth]{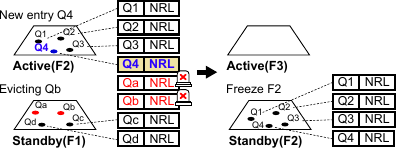}
    \caption[Query filter eviction.]{Query filter eviction.}
    \label{subfig:query-filt-dual}
    \vspace{-1.5mm}
\end{figure}

To avoid per-entry deletions while bounding the size of the query filter, 
\aker~ implements a dual-index design where it decouples insertions from evictions and reclaims space by dropping an entire index at epoch boundaries.
\aker~ maintains an \textit{active filter} $F_A$, \textit{standby filter} $F_S$, and a monotonically increasing epoch identifier.
During an epoch, $F_A$ admits new representative entries, whereas $F_S$ admits no inserts and is used to isolate evictions.
\aker~ advances the epoch when the number of evicted entries exceeds the number of live entries in the query filter.
At an epoch transition, the filters swap roles, as shown in Figure~\ref{subfig:query-filt-dual}.
Under FIFO, the entries are evicted only from $F_S$ (F1), and the evicted entries are marked as tombstones.
When the number of tombstones exceeds the number of live entries in the query filter,
\aker~halts inserts into $F_A$ (F2), seals it, and promotes it to become the next $F_S$.
Then, \aker~creates an empty index as next $F_A$ (F3) and drops the previous $F_S$ (F1), 
thereby reclaiming space without per-entry deletions or clean rebuilds.
During lookup, \aker~queries both $F_A$ and $F_S$ and selects the closest candidate cache entry.
Because evictions never occur in $F_A$, any candidate returned from $F_A$ is valid.
Consequently, \aker~ guarantees the return of valid cache entries, never returning tombstoned entries.

\vspace{-0.2em}
\section{Evaluation} \label{sec:evaluation}

We try to answer the following questions:
\begin{itemize}[leftmargin=*]
    \item Does the per-cached query similarity threshold maintain stable accuracy across different request distributions compared to single-threshold methods? (Section~\ref{subsec:eval-cache-acc})
    \item Can \aker~ reduce read amplification compared to a traversal-centric cache? (Section~\ref{subsec:eval-cache-eff})
    \item Can \aker~ sustain high performance without degrading recall under cache refreshes? (Section~\ref{subsec:eval-cache-ref})
\end{itemize} 

\subsection{Experiment Setup} \label{subsec:exp-setup}

We integrate \aker~ into pgvector~\cite{pgvector} HNSW (v0.8.0), a PostgreSQL~\cite{postgresql} extension, as our primary ANNS system.
pgvector provides a 
page-based traversal-centric cache baseline via PostgreSQL shared buffer. 
We run all experiments on a server equipped with a four-socket Intel Xeon Platinum 8276 CPU, $512$\,GB RAM, and a $512$\,GB NVMe SSD, running Ubuntu 24.04.
The client and PostgreSQL execute on the same machine.
The client issues requests from a single thread.
Unless otherwise stated, the default \aker~ uses $(\alpha_T, \alpha_L) = (0.25, 1.1)$, 
a query filter of m = 4, $efConstruction$ = 8, $efSearch$ = 16, and no deletion reserve ($\Delta = 0$). 
Section ~\ref{subsec:eval-cache-ref} evaluates variants 
derived from the default configuration.

PostgreSQL caches $8$KB table and index pages in its shared buffer~\cite{postgresql}.
pgvector stores page-organized vectors and indexes as PostgreSQL tuples.
These pages are also cached via the shared buffer, which cannot be disabled manually.
To reduce the impact of buffers, we set the size to the minimum $128$KB when evaluating approximate caches.
pgvector can benefit from both shared buffers and OS cache ~\cite{postgresqlresource}.
Thus, we choose two settings for search workloads: a $16$\,GB \textit{memory-constrained deployment} via container, 
and a \textit{host deployment} where pgvector runs directly on the host.
The memory-constrained deployment limits the OS caching support.

We also evaluate whether \aker~ improves searches on a highly optimized disk-based index, DiskANN ~\cite{jayaram2019diskann}.
We add \aker~ as a front-end caching layer for DiskANN.
DiskANN already manages its own memory-resident index, 
so \aker~ forwards a query to DiskANN only on a cache miss.
The benchmark runs with a single thread.

\smallskip
\noindent
\textbf{Datasets.}
We evaluate two application scenarios -- web search and RAG -- using SPACEV~\cite{spacev} and SPHERE~\cite{shen2025hermes} as summarized in Table~\ref{tab:build-confs}.
For RAG queries, we use TriviaQA~\cite{joshi2017triviaqa}.
For the search workload, we use $10$M-scale datasets; for the refresh stress test, we use a $1$M-scale dataset.
We use a smaller scale in the refresh test to increase the likelihood that newly inserted vectors appear in the top-$k$, thereby stressing the refresh behavior.
We build all indexes using Euclidean distance.
For each workload, we set search parameters to achieve $0.9$ recall at $k{=}10$ without approximate caches.

\begin{table}[h]
\normalsize
\vspace{-2mm}
\caption{Build configurations and data sizes}
\resizebox{\linewidth}{!}{%
\begin{tabular}{
    >{\centering\arraybackslash}p{1.33cm}
    >{\centering\arraybackslash}p{0.28\linewidth}
    >{\centering\arraybackslash}p{0.28\linewidth}
    >{\centering\arraybackslash}p{0.28\linewidth}
    }
    \toprule
    Dataset & SPACEV(10M) & SPHERE(10M) & SPACEV(1M) \\
        \midrule
    Dimension & 100 & 768 & 100 \\
        \midrule
    \makecell{pgvector} &
        \makecell{\texttt{m=16}, \texttt{efCon=128},\\Vector $4.6$\,GB,\\Index $7.4$\,GB} &
        \makecell{\texttt{m=32}, \texttt{efCon=256},\\Vector $41.0$\,GB,\\Index $40.9$\,GB} &
        \makecell{\texttt{m=16}, \texttt{efCon=64},\\Vector $477.9$\,MB,\\Index $781.8$\,MB} \\
        \midrule
    \makecell{DiskANN} &
        \makecell{\texttt{R=64}, \texttt{L=100},\\On-disk $3.5$\,GB,\\In-mem $1.1$\,GB} &
        \makecell{\texttt{R=96}, \texttt{L=100},\\On-disk $39.0$\,GB,\\In-mem $5.1$\,GB} &
        \makecell{-} \\
    \bottomrule
\end{tabular}%
}
\label{tab:build-confs}
\vspace{-2mm}
\end{table}

\begin{figure}[t]
    \centering
    \includegraphics[width=0.91\linewidth]{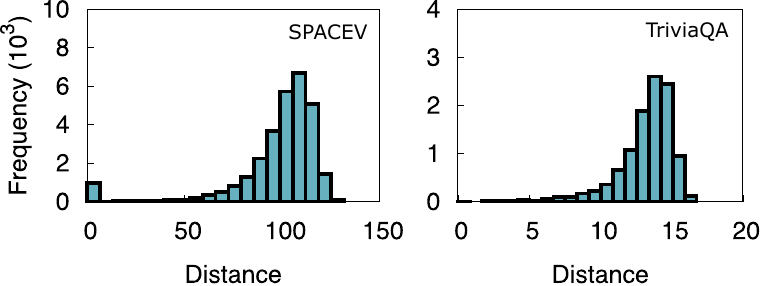}
    \caption[Query nearest-neighbor (top-1) distance distribution]{
    \textbf{Nearest-neighbor (top-1) distance distribution in a query set.}
    Smaller distances indicate that queries are more similar to one another.
    }
    \label{fig:rev-query-nnd-distribution}
    \vspace{-2mm}
\end{figure}

\noindent
\textbf{Queries.}
We first examine the two released query sets: SPACEV and encoded TriviaQA.
Figure~\ref{fig:rev-query-nnd-distribution} shows, for each query, the distance to its nearest neighbor in the released query set.
The distribution characterizes inter-query relationships.
The released queries show that zero distance is rare, indicating that most queries are distinct.
Second, both dataset contain many queries with nearby neighbors.
The released query sets contain semantic neighborhoods whose top-$k$ results can overlap, making approximate reuse meaningful.
However, real-world workloads are often skewed~\cite{ycsbc, bergman2025leveragingmid, adamic2002zipf}. 
Public datasets provide query collections for measuring retrieval quality and are widely used in ANNS benchmarks~\cite{shen2025hermes, neurips21comp, neurips21comp, bigann, spacev}. 
For example, BIGANN~\cite{bigann} releases 10K benchmark query vectors, 
and SPACEV~\cite{spacev} provides query descriptors from Bing web searches. 
TriviaQA~\cite{joshi2017triviaqa} comprises question-answer pairs collected from websites. 
These datasets are released as sampled query collections and therefore do not preserve request pattern information.

Because real traces are unavailable, 
we synthesize a workload called \textit{simZipf} that jointly models query similarity and request skew. 
We construct simZipf in two stages.
First, for each base query, 
we build a similar-query group containing 50 queries.
We take two different synthesis methods because each dataset exposes different forms of query information.
For SPACEV, which provides only a query set $Q$ of 29K vectors, 
we construct nearby queries directly in the vector space via linear interpolation.
For each base vector $q_i \in Q$, we sample another vector $p \in Q$ and generate
$q_i' = (1-\lambda)q_i + \lambda p$, where $\lambda$ is sampled uniformly from $(0, 0.5)$.
Repeating this procedure with 50 sampled choices of $p$ yields the similar-query group.
For TriviaQA, which provides 10K raw natural-language questions, 
we generate 50 semantically similar questions using Mistral-7B~\cite{chaplot2023albert}.
Then, we encode the generated questions with the BGE-large encoder~\cite{xiao2024c} to form the similar-query group.
Next, we generate a skewed request sequence over the similar-query groups using a Zipfian distribution~\cite{yang2021large, cooper2010benchmarking, bergman2025leveragingmid, adamic2002zipf}.
We use YCSB-C~\cite{ycsbc} to generate a $100$K Zipfian key sequence $(k_1, k_2, \dots)$, 
and map each similar-query group to the key with the same popularity rank.
For each key access, 
we uniformly sample one query from the corresponding similar-query group.

simZipf aims to simulate temporal locality while preserving semantic neighborhoods.
We use three different skew settings ranging from low to high (Zipfian exponents: $0.30$, $0.60$, and $0.99$) to simulate different combinations of semantic and temporal locality.
Approximate caches rely on two locality dimensions: whether queries appear nearby in the vector space, and how strongly requests concentrate over the neighborhoods.
To clarify these two dimensions, we analyze the simZipf traces using two complementary statistics:
(1) nearest-neighbor distance distribution and
(2) repeated-query distribution.
The first captures whether the generated traces remain concentrated within semantic neighborhoods.
The second captures how much of the locality stems from repeated identical queries compared to the semantically similar queries.

\begin{figure}[t]
    \centering
    \begin{subfigure}[b]{0.93\linewidth}
        \centering
        \includegraphics[width=\linewidth]{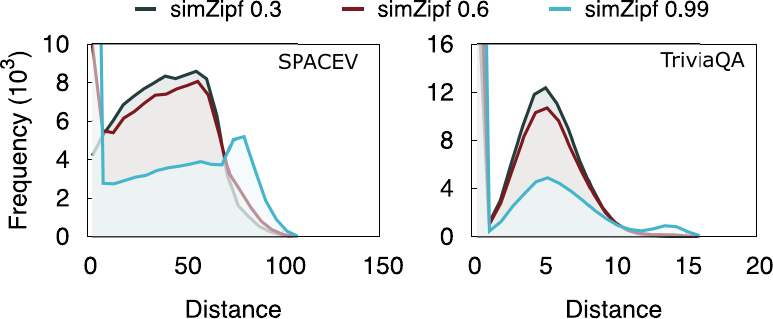}
        \vspace{-0.6em}
        \caption[Generated nearest-neighbor distance distribution]{
            \textbf{Generated nearest-neighbor distance distribution.}}
        \label{fig:rev-gen-nnd-distribution}
        \vspace{-0.2em}
    \end{subfigure}
    \begin{subfigure}[b]{0.93\linewidth}
        \centering
        \includegraphics[width=\linewidth]{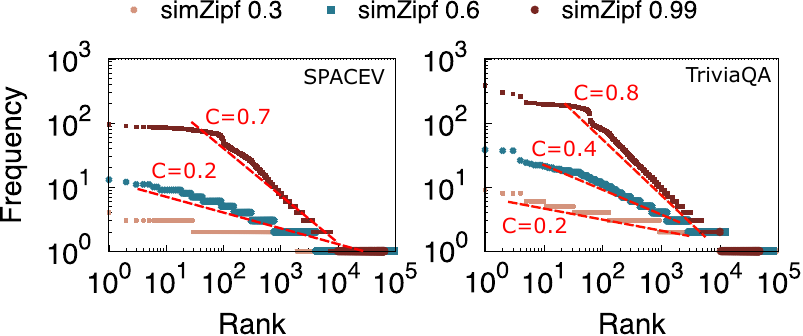}
        \caption[Equal-query distribution]{Equal-query distribution.}
        \label{fig:rev-exact-rep}
    \end{subfigure}
    \label{fig:rev-distribution}
    \vspace{-0.6em}
    \caption[Similarity-popularity of simZipf queries]{
        \textbf{Similarity-popularity of simZipf queries.}}
    \vspace{-1.5mm}
\end{figure}

Figure~\ref{fig:rev-gen-nnd-distribution} shows that as the simZipf skew weakens, 
the generated traces spread toward larger nearest-neighbor distances.
The two datasets exhibit different distance profiles.
TriviaQA shows a bell-shaped distribution in mid-range distances (roughly $2$ to $10$).
SPACEV shows a flatter profile (roughly $10$ to $70$) because its similar-query groups are generated directly in vector space.
Despite these differences, both datasets exhibit the same qualitative trend: 
lower skew shifts more requests toward more distant semantic neighbors.
This makes approximate reuse harder because fewer requests concentrate in very close neighborhoods, forcing reuse to rely on weaker semantic locality.
In this sense, lower-skew traces are harsher for approximate caching, much as workloads with weaker temporal locality are harsher for conventional caches.

We also analyze how often identical queries repeat in the generated traces.
This analysis is necessary because if the workload were dominated by a very small number of repeated queries, 
a large fraction of the cache benefit could be attributed to equality matches.
Figure~\ref{fig:rev-exact-rep} shows that stronger simZipf skew does not cause the traces to be dominated by repeated queries.
The repeated-query distribution has a steeper cutoff and faster decay than would result from applying the same Zipfian skew directly to individual queries.

\smallskip
\noindent
\textbf{Baselines.}
We compare \aker~ against Potluck and \textsc{Proximity}, which are the latest approximate caches.
We also compare against PostgreSQL's default traversal-centric cache, the shared buffer.
For a fair comparison, we implement \textsc{Proximity} and Potluck within \aker~ by substituting only the threshold control module and keeping the rest of the system unchanged.
All baselines share \aker's neighbor object pool, \texttt{NRL} layout, and eviction policy.
The differences reflect only the cache-hit rule.
In both baselines, alias entries are disabled as they are \aker-specific optimizations.

\begin{itemize}[leftmargin=*]
    \item \textbf{Potluck+}~\cite{guo2018potluck}.
        We implement Potluck's threshold-tuning mechanism.
        Following the Potluck algorithm, upon cache-entry insertion, we test set equivalence using the hashed vectors stored in \texttt{NRL}.
        At lookup time, Potluck performs dropout: with probability $p$, it forces a cache miss.
        We use Potluck's default setting of $p{=}0.10$ and add a more aggressive setting of $p{=}0.25$.
        
    \item \textbf{\textsc{Proximity}+}~\cite{bergman2025leveragingmid}.
        We use a strict threshold for high recall.
        We measure all inter-query distances between the synthesized queries 
        and use the 1st-percentile value as the fixed threshold.
\end{itemize}

\subsection{Cache Accuracy} \label{subsec:eval-cache-acc}

\begin{figure}[t]
    \centering
    \includegraphics[width=1\linewidth]{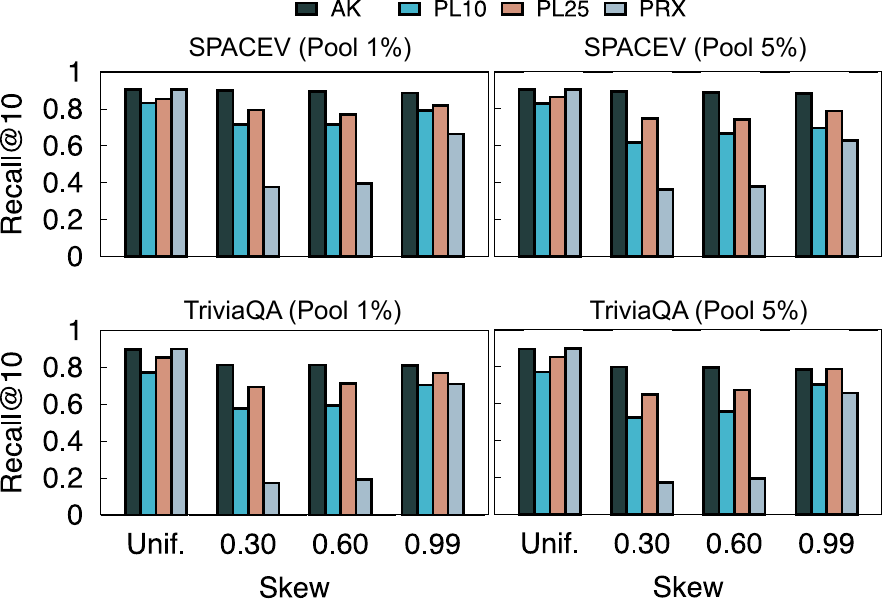}
    \caption{
        \textbf{Recall.}
        \aker~(AK) maintains higher recall across skew levels than Potluck+ (PL) and \textsc{Proximity}+ (PRX).
    }
    \label{fig:recall-stab}
\end{figure}

\begin{figure}[t]
    \centering
    \includegraphics[width=1\linewidth]{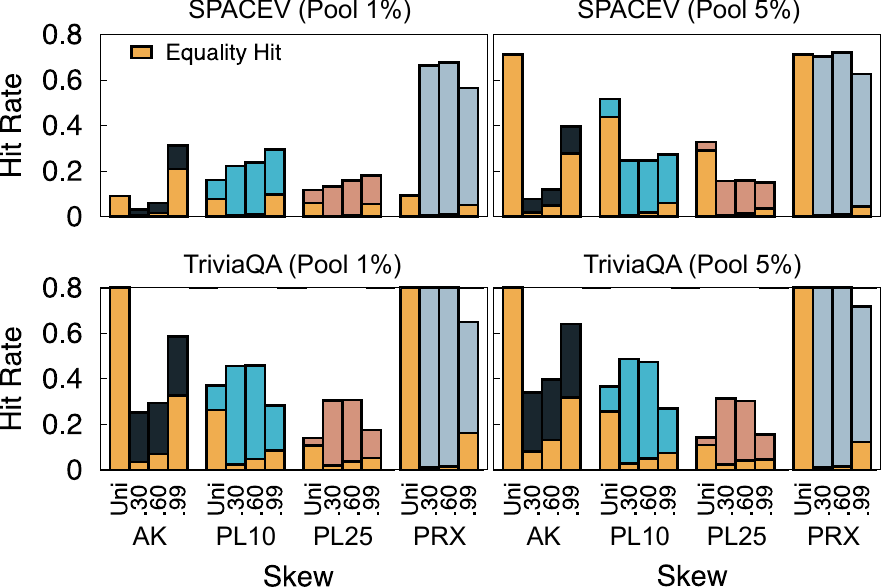}
    \caption{
    \textbf{Hit ratios.}
    Each stacked bar shows the total cache hit ratio.
    The yellow segment denotes equality hits, and the remaining segment denotes approximate hits.
    }
    \label{fig:hitrate-breakdowns}
\end{figure}

We evaluate all cache methods at $k=10$.
This is a strict criterion, as each single missing ground truth neighbor reduces recall by $0.1$.
We target at least $0.8$ recall for \aker, corresponding to fewer than one missed true neighbor per query.
We set the neighbor object pool to store up to $1\%$ and $5\%$ of the dataset vectors.
\aker~ shows consistent trends in recall and hit ratios across pgvector and DiskANN; thus we provide results for pgvector in Figure~\ref{fig:recall-stab} and~\ref{fig:hitrate-breakdowns}.

\smallskip
\noindent
\textbf{Accuracy.}
Across varying skewness, 
\aker~ achieves up to $52.3$ percentage points (pp) and $64.0$ pp higher recall than the single similarity threshold methods on SPACEV and TriviaQA, respectively.
In all settings, \aker~ misses fewer than one correct result on average and maintains recall above $0.80$.
The single similarity thresholds become overly permissive, producing more approximate hits.
The low recall persists even when \textsc{Proximity}+ (PRX) uses a conservative setting of 1\% inter-query distance, indicating that a strict constraint is insufficient to preserve accuracy.

Potluck+ (PL) achieves higher recall than PRX by forcing pgvector searches.
Higher $p$ leads to a higher recall.
With $p=0.1$ (PL10), the maximum recall reaches $0.79$ and $0.71$ on each dataset, still missing more than one correct item.
Higher recall is achieved by more frequently invoking the ANNS.
Increasing $p$ to $0.25$ (PL25) improves recall compared to PL10, but both still remain worse than \aker.
Under low skew, query distributions become diverse.
A single threshold fails to capture query-specific density.
Thus, as skew decreases, the recall gap between \aker~ and other methods widens.

\smallskip
\noindent
\textbf{Cache hit ratios.}
Figure~\ref{fig:hitrate-breakdowns} shows the hit ratios under varying skewness levels ($x$-axis) and cache sizes. 
As skew increases, near-duplicate queries recur more frequently; all caches show increased total hit ratios. 
PRX achieves high hit ratios across all configurations, but mostly via approximate hits, which substantially reduce recall.
PL also shows a higher approximate hit ratio than \aker.
Although a larger $p$ triggers more ANNS, PL's reliance on a single similarity threshold produces high approximate hits.
In contrast, \aker~ adapts to workload skewness. 
As skew grows, \aker~ relies less on approximate hits.
\aker~ shows higher recall because multiple thresholds allow approximate hits only when they are locally safe.

The uniform distribution further highlights the difference in the threshold control.
As queries recur, equality hits dominate in \aker~ and the recall is preserved.
In PL, the equality hits gradually loosen its global threshold and allows farther approximate matches.

\begin{figure*}[t]
    \centering
    \includegraphics[width=0.95\linewidth]{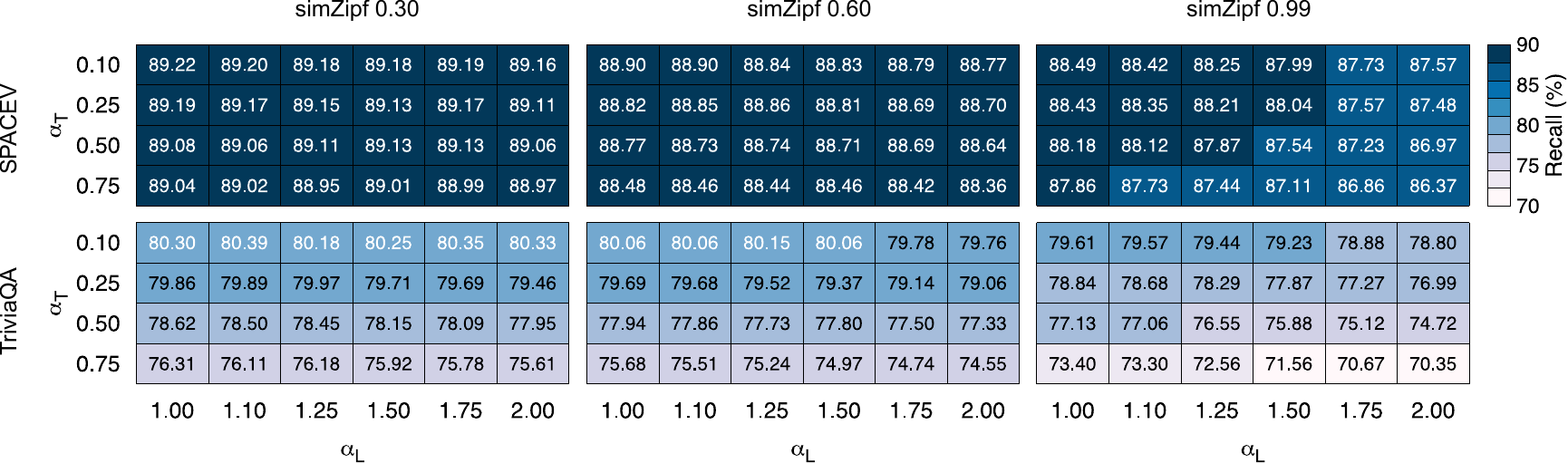}
    \caption[Sensitivity study: Recall]{
        \textbf{\aker~ Sensitivity study: Recall.}
        Each cell shows recall (\%) for one parameter 
        pair $(\alpha_T, \alpha_L)$.
        }
    \label{fig:rev-sensitivity-recall}
    \vspace{-3mm}
\end{figure*}

\begin{figure*}[t]
    \centering
    \begin{subfigure}[b]{\linewidth}
        \centering
        \includegraphics[width=0.95\linewidth]{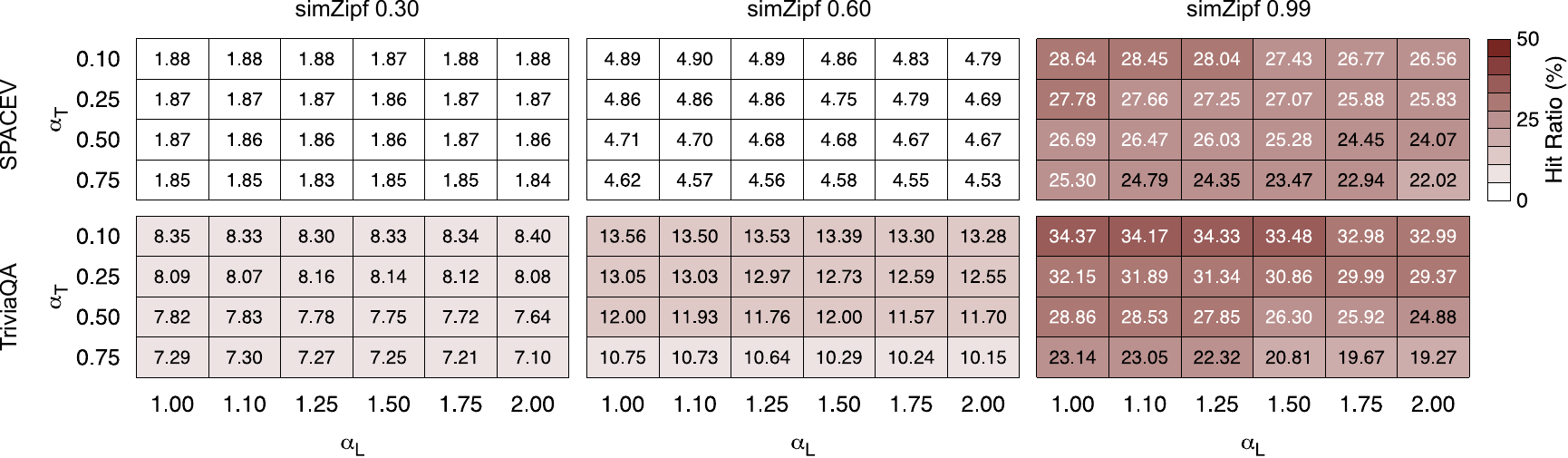}
        \caption[Equality hit ratio]{
            \textbf{Equality hit ratio}.
        }
        \label{fig:rev-sensitivity-exact-hr}
    \end{subfigure}
    \begin{subfigure}[b]{\linewidth}
        \centering
        \includegraphics[width=0.95\linewidth]{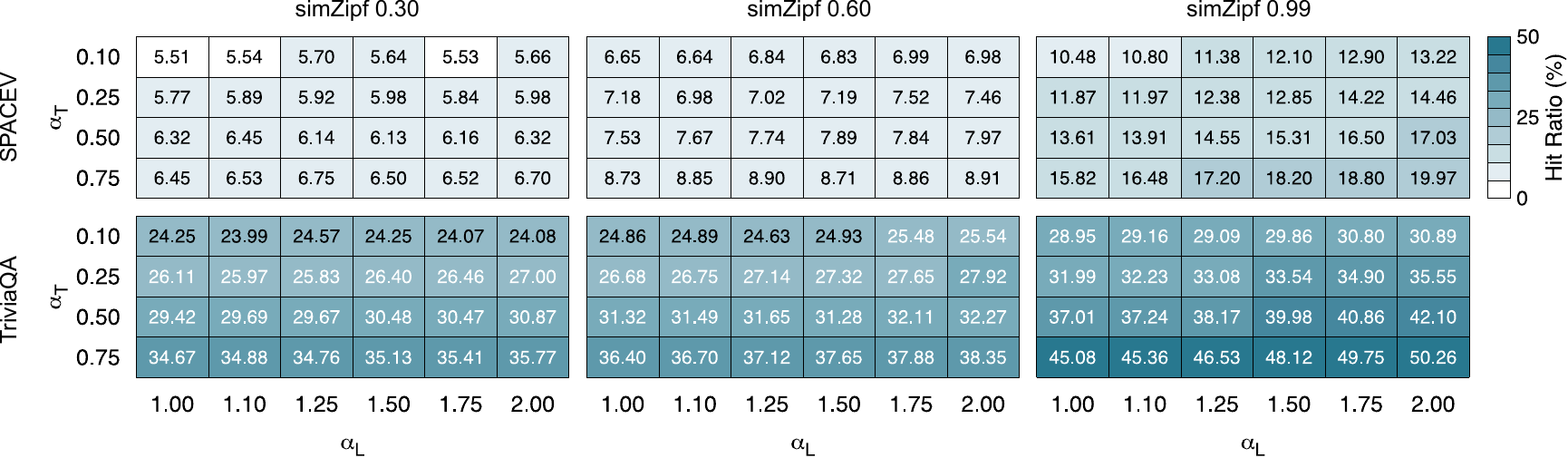}
        \caption[Equality hit ratio]{
            \textbf{Approximate hit ratio}.
        }
        \label{fig:rev-sensitivity-approx-hr}
        \vspace{-1.5mm}
    \end{subfigure}
    
    \caption[\aker~Sensitivity study]{
        \textbf{\aker~ Sensitivity study: Hit ratio}.
        Each cell shows hit ratio (\%) for one parameter 
        pair $(\alpha_T, \alpha_L)$.
    }
    \label{fig:rev-sensitivity-study}
\end{figure*}

\smallskip
\noindent
\textbf{Sensitivity study}.
We empirically show how \aker~ preserves accuracy under different parameter choices.
We sweep $\alpha_T$ and $\alpha_L$ at 5\% pool size on SPACEV and TriviaQA under simZipf 0.3, 0.6, and 0.99.
The sweep shows two consistent trends.
First, large $\alpha_L$ reduces recall and equality hit ratios.
A larger $\alpha_L$ expands $\tau_q$ more aggressively, 
which makes later cache lookups more likely to be served by distant approximate hits.
Second, large $\alpha_T$ also reduces recall and equality hit ratios.
Because $\alpha_T$ controls contraction speed after approximate hits, a larger $\alpha_T$ keeps $\tau_q$ loose for longer.
As a result, more approximate hits are allowed.

The sensitivity remains limited under low skew.
Under simZipf $0.3$, the workload shifts toward more diverse mid-range queries, reducing opportunities for safe reuse.
Even in this setting, however, \aker~ prevents large recall loss across the parameter sweep.
On SPACEV, recall changes only marginally across the tested parameter pairs.
On TriviaQA, the recall spread is $4.8$ percentage points (pp).
However, the sensitivity becomes larger under high skew.
For example, on TriviaQA, the recall spread increases from $4.8$ pp at simZipf $0.3$ to $9.2$ pp at simZipf $0.99$.

Figure~\ref{fig:rev-sensitivity-exact-hr} and Figure~\ref{fig:rev-sensitivity-approx-hr} explain why the gap widens.
As either $\alpha_T$ or $\alpha_L$ increases, 
the share of equality hits decreases while the share of approximate hits increases.
When a query is first served by an approximate hit, 
\aker~ creates an alias entry for that query and links it to the previously served entry.
Subsequent repetitions of those equal queries are then served through the alias path.
Looser parameter settings therefore admit more initial approximate hits, 
create more alias entries, and divert more future repetitions away from the equality hit path.
Thus, the equality hit ratio falls even though more requests are being served from cache.
The effect is clearest on TriviaQA with simZipf $0.99$.
At $\alpha_L=1.10$, increasing $\alpha_T$ lowers the equality hit ratio from $34.2$\% to $23.1$\% while increasing the approximate hit ratio from $29.7$\% to $45.4$\%.
At $\alpha_T=0.25$, increasing $\alpha_L$ from $1.10$ to $2.00$ lowers the equality hit ratio from $31.9$\% to $29.4$\% and increases the approximate hit ratio from $32.2$\% to $35.6$\%.
Across both figures, $\alpha_T$ is the more influential parameter.
Because $\alpha_T$ directly controls how aggressively \aker~ contracts the threshold, increasing $\alpha_T$ produces a larger shift toward approximate reuse and a larger recall drop than increasing $\alpha_L$.

\vspace{-2mm}
\subsection{Cache Efficiency} \label{subsec:eval-cache-eff}

\begin{figure}[t]
    \centering
    \begin{subfigure}[b]{0.52\linewidth}
        \centering
        \includegraphics[width=\linewidth]{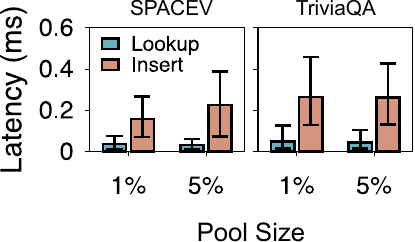}
        \caption[\aker~ latency]{\aker~ latency}
        \label{fig:rev-cache-internals}
    \end{subfigure}
    \begin{subfigure}[b]{0.47\linewidth}
        \centering
        \includegraphics[width=\linewidth]{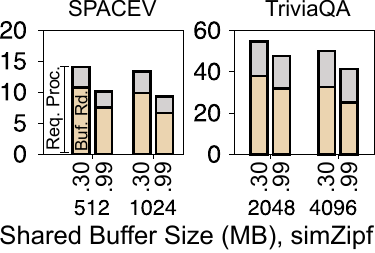}
        \caption[pgvector latency]{pgvector latency}
        \label{fig:rev-pgvector-internals}
    \end{subfigure}
    \caption{
    \textbf{\aker~ (lookup and cache entry insert) and pgvector latency.}
    Each bar shows the average, and error bar shows 1st- and 99th-percentile latencies.
    }
    \label{fig:rev-internals}
    \vspace{-2mm}
\end{figure}

\noindent
\textbf{Operational overhead}.
Figure~\ref{fig:rev-internals} shows that \aker~ adds little overhead compared to pgvector processing latency.
The average cache lookup latency remains below $0.1$\,ms.
A cache miss adds less than $1$\,ms, from cache lookup and cache entry insertion combined.

\smallskip
\noindent
\textbf{Read amplification.}
We observe read amplification by measuring the average number of PostgreSQL $8$\,KB pages read per query from PostgreSQL logs.
We compare \aker~(AK) against pgvector's traversal-centric cache, PostgreSQL shared buffer (PG).
For \aker, we set the maximum neighbor object pool size to 1\%, 2\%, and 5\%.
For the shared buffer, we set the size from $128$\,MB to $8$\,GB.

\begin{figure}[b]
    \centering
    \vspace{0.3em}
    \includegraphics[width=1\linewidth]{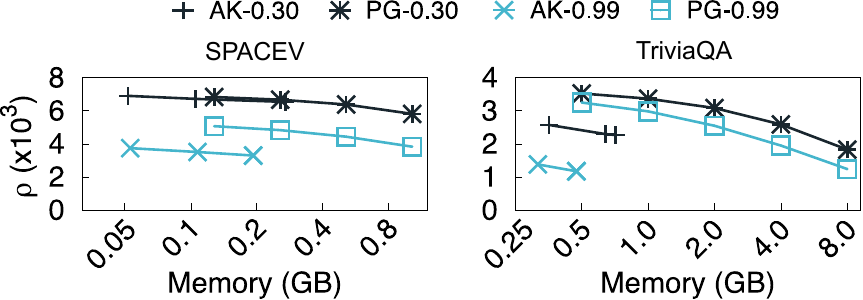}
    \caption{
        \textbf{Read amplification.}
        Numeric labels show the skewness. 
        We omit simZipf 0.60 as it follows the same trend as simZipf 0.30.
        The memory use of the query filter is approximated as:
        (1) \(R\) query vectors of cache entries (dimension \(d\), \texttt{float32}) are \(R \cdot d \cdot 4\) bytes,
        (2) maximum number of neighbors \(m = 4\), storing \(m\) 8-byte identifiers per entry are \(R \cdot m \cdot 8\) bytes.
    }
    \label{fig:ra-line}
\end{figure}

Figure~\ref{fig:ra-line} shows that \aker~reduces read amplification, $\rho$, on both datasets.
Under high skew (0.99), \aker~reduces $\rho$ to $0.7\times$ and $0.4\times$ of pgvector on SPACEV and TriviaQA, while using only $0.4\times$ and $0.5\times$ of shared buffer memory, respectively.
Under low skew (0.30), \aker~still achieves $1.0\times$ and $0.7\times$ $\rho$ in both datasets while using $0.8\times$ and $0.6\times$ of pgvector's memory.
As ANNS requires large reads, each query can still read many pages when the working set exceeds the buffer.
In \aker, page reads occur only on cache misses.

\begin{figure}[t]
    \centering
    \begin{subfigure}[b]{1\linewidth}
        \centering
        \includegraphics[width=1\linewidth]{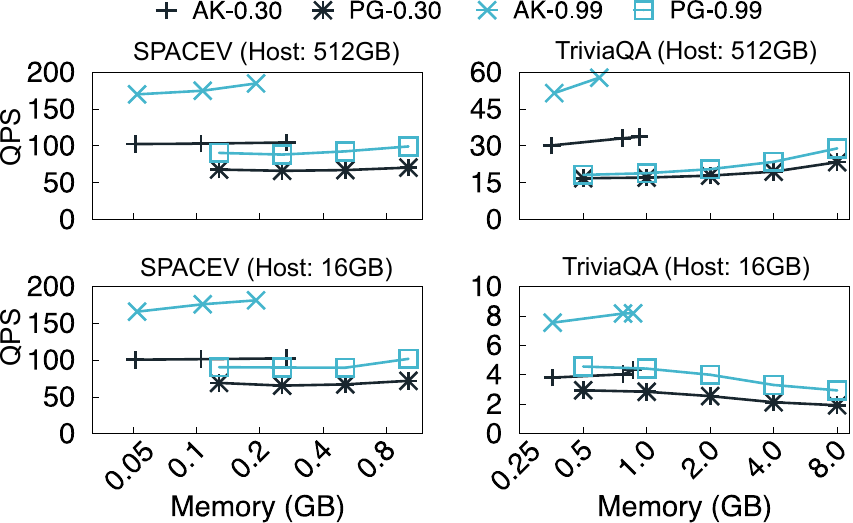}
        \caption{
            \textbf{QPS of pgvector.}
        }
        \label{fig:qps-line}
    \end{subfigure}
    \begin{subfigure}[b]{1\linewidth}
        \centering
        \includegraphics[width=1\linewidth]{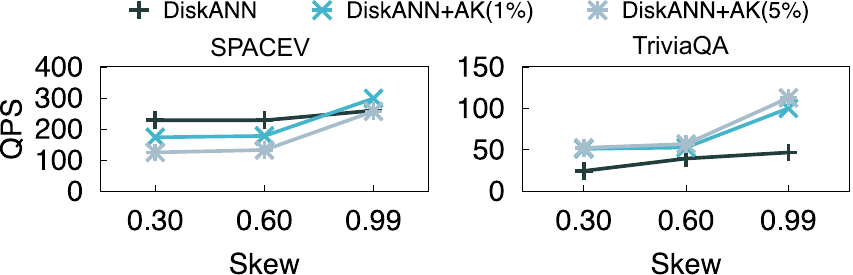}
        \caption{
            \textbf{QPS of DiskANN.} The additional memory footprint of \aker~ over DiskANN remains small, 
            reaching at most 26.4\,MB on SPACEV and 275.6\,MB on TriviaQA, 
            which corresponds to $2.5$\% and $5.4$\% of DiskANN's pinned in-memory index, respectively.
        }
        \label{fig:qps-line-diskann}
        \vspace{-1mm}
    \end{subfigure}
    \caption{\textbf{QPS.} Figure ~\ref{fig:qps-line} and Figure ~\ref{fig:qps-line-diskann} follow the same approximation and plotting convention as in Figure ~\ref{fig:ra-line}.
    }
    \vspace{-2mm}
    \label{fig:rev-aker-internals}
\end{figure}

\smallskip
\noindent
\textbf{QPS}.
Figure~\ref{fig:qps-line} reports QPS under two pgvector deployment settings. 
On SPACEV, the $16$\,GB limit shows little difference in QPS as the data size fits the memory budget. 
On TriviaQA, in contrast, QPS decreases 
because the data is larger than the memory budget.
In all settings, 
\aker~ achieves up to $2\times$ and $3\times$ QPS while using $0.8\times$ and $0.6\times$ the memory on SPACEV and TriviaQA, 
respectively. 

Despite DiskANN’s strong baseline efficiency, \aker~ provides improvements, 
with benefits varying across datasets and workload skew.
On 100-dimensional SPACEV, DiskANN search is inexpensive, 
achieving an average query latency of $3.7$\,ms, 
with 70\% spent on reads and 30\% on index traversal.
Under lower-skew workloads, the hit rate is not high enough to amortize \aker's lookup cost shown in Figure ~\ref{fig:rev-internals}.
Thus, QPS improves only at simZipf 0.99.
On 768-dimensional TriviaQA, DiskANN search is substantially more expensive:
the average query latency rises to $21.1$\,ms, with 31\% spent on reads and 67\% on index traversal.
Although this latency is lower than that of pgvector, 
the higher backend search cost makes bypassing the full search path more beneficial.
Thus, \aker~ achieves up to $2.4\times$ higher QPS across all skews on TriviaQA.

\begin{figure}[b]
    \centering
    \begin{subfigure}[b]{0.48\linewidth}
        \centering
        \includegraphics[width=\linewidth]{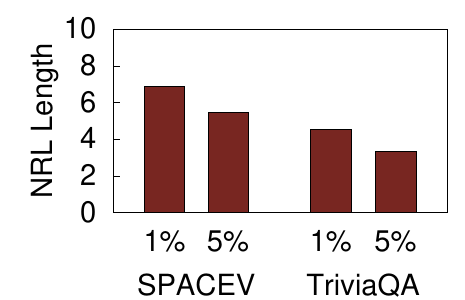}
        \caption[NRL length]{
            \textbf{\texttt{NRL} length}
            }
        \label{subfig:centry-breakdown}
    \end{subfigure}
    \begin{subfigure}[b]{0.48\linewidth}
        \centering
        \includegraphics[width=\linewidth]{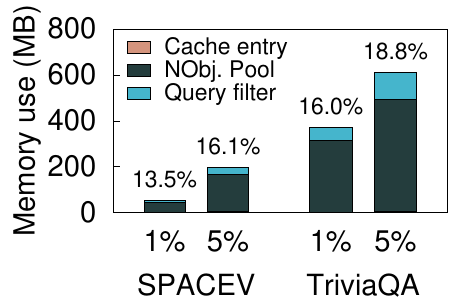}
        \caption[Query filter overhead]{
            Query filter overhead
        }
        \label{subfig:mem-breakdown}
    \end{subfigure}
    \caption{
        \textbf{Memory efficiency.}
        Figure~\ref{subfig:mem-breakdown} shows the memory usage, following the same approximation of Figure ~\ref{fig:qps-line}.
    }
    \label{fig:breakdowns}
\end{figure}

\smallskip
\noindent
\textbf{Memory efficiency.}
Figure~\ref{fig:breakdowns} shows \texttt{NRL} lengths and the query filter's share of total memory in simZipf 0.99.
An \texttt{NRL} length is the ratio of the number of neighbor objects to the number of cache entries.
The \texttt{NRL} length below $k=10$ indicates neighbor objects are shared across entries.
Compared to a naive design that allocates $k$ neighbor objects per entry, 
\aker~ reduces the number of neighbor objects needed by $2.9$-$6.1\times$.
Figure~\ref{subfig:mem-breakdown} shows that the number of cache entries brings minor overhead, 
with the query filter accounting for only $13.5$-$18.8$\% of total memory.

\subsection{Cache Refresh} \label{subsec:eval-cache-ref}

To invoke cache refresh, we first make the cache dirty, then run searches on the \aker-attached pgvector.
Using a 1\% neighbor object pool, we first warm up the cache by running simZipf 0.99.
Next, we insert an additional 5\% of vectors to fill the insert log from another split of SPACEV~\cite{xu2023spfresh}.
We then randomly invalidate neighbor objects to simulate deletions.
We use deletion levels of 5\%, 10\% and 25\%.
Finally, the same search queries are re-executed.

\begin{figure}[t]
    \centering
    \begin{subfigure}[b]{0.51\linewidth}
        \centering
        \includegraphics[width=\linewidth]{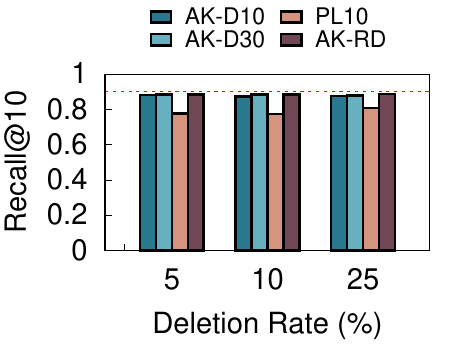}
        \caption[Recall]{
            Recall
        }
        \vspace{-2mm}
        \label{fig:stress-recall}
    \end{subfigure}
    \begin{subfigure}[b]{0.48\linewidth}
        \centering
        \includegraphics[width=\linewidth]{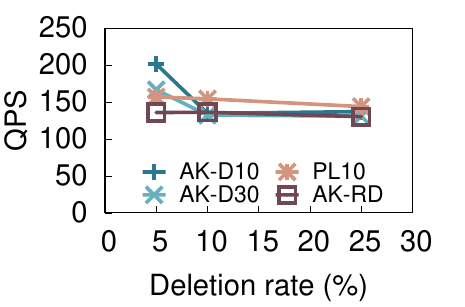}
        \caption[QPS]{
            QPS
        }
        \vspace{-2mm}
        \label{fig:stress-qps}
    \end{subfigure}
    \caption{
        \textbf{Refresh stress-test under updates.}
        The red line is the recall when no approximate cache is attached.
        \aker~ (AK) maintains high recall and QPS compared to Potluck+ (PL) and the refresh-disabled variant of \aker~(RD).
    }
    \label{fig:stress-test-perf}
    \vspace{-2mm}
\end{figure}

For \aker, we set $\Delta$ to $10$ (AK-D10) and $30$ (AK-D30), the risk threshold to $0.3$, and the slow path batch size $B$ to $16$.
We compare against Potluck+ with probability $p=0.10$ (PL10).
We include a no-refresh ablation with $\Delta=0$, AK-RD, in which refresh mechanisms are disabled in \aker.
For PL10 and AK-RD, no refresh is performed.
Under deletions, if a cache entry cannot maintain $k$ valid results, it is treated as a miss to avoid correctness violations (P\ref{prop:corr}).

\smallskip
\noindent
\textbf{Accuracy and QPS.}
AK-D10 and AK-D30 incur at most $2$ pp recall drop across all deletion rates.
AK-RD shows the highest recall because deletions invalidate many entries and trigger the ANNS more frequently.
In Table~\ref{tab:stress-test-hitrate}, AK-RD shows a low cache hit ratio of $6.4$\% due to the large number of invalidated cache entries.
In contrast, PL10 suffers from a large recall drop of $12.3$ pp, even at low deletion rates.
At a deletion rate of 5\%, AK-D10 achieves the highest QPS, outperforming PL10 by 1.3$\times$ and AK-RD by 1.5$\times$. 

\begin{table}[b]
\caption{Number of entries and hit ratio at 5\% deletion rate.}
\resizebox{\linewidth}{!}{%
\begin{tabular}{
    >{\centering\arraybackslash}p{1.4cm}
    >{\centering\arraybackslash}p{0.20\linewidth}
    >{\centering\arraybackslash}p{0.20\linewidth}
    >{\centering\arraybackslash}p{0.20\linewidth}
    >{\centering\arraybackslash}p{0.20\linewidth}
}
    \toprule
     & AK-D10 & AK-D30 & PL10 & AK-RD \\
    \midrule
    \# of entries & $625$ & $315$ & $1,170$ & $1,284$ \\
    \midrule
    Hit ratio & 12.52\% & 9.71\% & 16.64\% & 6.39\% \\
    \bottomrule 
\end{tabular}%
}
\label{tab:stress-test-hitrate}
\end{table}

\smallskip
\noindent
\textbf{Selection of $\Delta$.}
Under a fixed pool budget, 
increasing $\Delta$ improves per-entry tolerance to deletions 
(more reserves per cached query), 
but each cache entry consumes more neighbor objects.
Thus, the number of cache entries decreases.
Table~\ref{tab:stress-test-hitrate} illustrates the effect.
AK-D10 maintains $2\times$ more cache entries with a higher hit ratio than AK-D30.
Because AK-RD cannot recover cache entries, it shows the lowest hit ratio and QPS, even though it maintains $4\times$ more cache entries than AK-D30.
Overall, AK-D10 achieves the best deletion tolerance--recall balance, 
including the most aggressive stress level up to 25\% deletions. 
This suggests that a small $\Delta$ is sufficient to tolerate the deletion pressure.

\begin{figure}[t]
    \centering
    \includegraphics[width=0.97\linewidth]{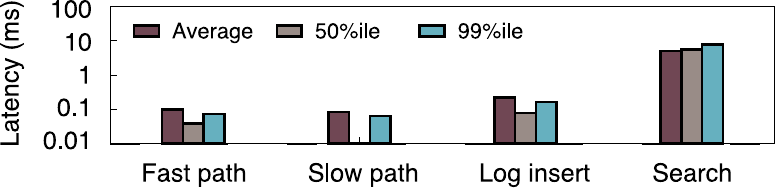}
    \caption[Cache refresh overhead]{
        \textbf{Cache refresh overhead (AK-D10).}
    }
    \label{fig:rev-refresh-lat-internals}
    \vspace{-2mm}
\end{figure}

\smallskip
\noindent
\textbf{Refresh overhead.}
Figure~\ref{fig:rev-refresh-lat-internals} reports the refresh latency in AK-D10 and compares it against the end-to-end search latency of pgvector.
The log insert latency includes both the fast and slow paths.
Overall, the refresh overhead remains small, never exceeding $0.25$\,ms.
On average, \aker~adds only $0.22$\,ms, which is about $4.4$\% of the $5$\,ms search latency.
The results indicate that \aker~introduces negligible refresh overhead relative to query processing.

\vspace{-0.3em}
\section{Related Work} \label{sec:related-work}

\noindent
\textbf{ANN index optimizations.}
These studies propose optimized ANN index designs that reduce disk I/O and memory costs.
GoVector~\cite{zhou2025govector} proposes an index layout that places similar vectors on the adjacent disk pages to reduce I/O.
DiskANN~\cite{jayaram2019diskann} and SPANN~\cite{chen2021spann} propose a memory--disk hybrid index design to overcome extensive memory use.
DiskANN proposes the Vanama algorithm and a hybrid storage layout that keeps quantized vectors in-memory and full-precision vectors on-disk.
SPANN proposes a balanced clustering algorithm to reduce disk I/O overhead.
SPFresh~\cite{xu2023spfresh} extends SPANN to support incremental index builds while maintaining high accuracy under vector updates.
CXL-ANNS~\cite{jang2023cxl} and SmartANNS~\cite{tian2025towards} propose hardware-optimized index designs to accelerate index traversal.

\smallskip
\noindent
\textbf{Approximate caches.}
These studies introduce approximate caching across diverse domains.
Prior works have used approximate caches in systems for advertising~\cite{pandey2009nearest}, image retrieval~\cite{falchi2012similarity}, and scene understanding~\cite{selvam2025simcache}.
Potluck~\cite{guo2018potluck} proposed function-level caching to reduce cross-application computation.
RAGCache~\cite{jin2024ragcache} caches frequently referenced knowledge chunks to reduce end-to-end response time in RAG pipelines.
GPTCache~\cite{bang2023gptcache} and \textsc{Proximity}~\cite{bergman2025leveragingmid} leverage query similarity to reduce redundant document searches.
These systems do not address query-specific reuse control or cache refreshes, which are the key designs of \aker.

\vspace{-0.3em}
\section{Conclusion} \label{sec:conclusion}

We present \aker, 
an approximate cache for disk-based ANNS.
\aker~ targets two challenges in approximate caching: 
maintaining high recall and throughput, and designing a low-cost cache refresh mechanism.
First, \aker~ introduces per-query similarity thresholds that tailor hit decisions to each cached query based on observed hit behavior, preserving recall across varying neighborhood densities.
Second, \aker~ proposes del-consistency, which applies deletions eagerly while handling insertions lazily via $\Delta$ reserves and risk scores that enable a low-overhead cache refresh pipeline.
We integrate \aker~into pgvector.
\aker~improves recall by up to $64$ percentage points over prior solutions and increases QPS by up to $3.2\times$ relative to PostgreSQL's shared buffer, while using only $0.6\times$ 
of the memory.

\begin{acks}
We thank the anonymous reviewers for their valuable comments and suggestions.
This research was supported by the IITP-MSIT (Institute for Information \& Communications Technology Planning \& Evaluation, Ministry of Science and ICT) (RS-2024-00436680 and RS-2024-00459026), Microsoft Research, and NRF (National Research Foundation of Korea) (RS-2025-00518369).
\end{acks}


\bibliographystyle{ACM-Reference-Format}
\bibliography{reference}


\end{document}